HIGHER SCHOOL OF ECONOMICS
NATIONAL RESEARCH UNIVERSITY

# NETWORK ANALYSIS OF INTERNATIONAL MIGRATION

The paper analyses international migration flows from the network perspective by the evaluation of centrality indices. In order to find the most influential countries in the international migration network classical centrality indices and new centrality indices are evaluated. New centrality indices consider short (SRIC) and long-range (LRIC) indirect interactions and the node attribute – population of the destination country. The model is applied to the annual data on international migration flows from 1970 to 2013 provided by United Nations Organization. The analysis is made for one year of each decade and indices' dynamics is described. It is shown that countries with huge migration flows are outlined by both classical and SRIC, LRIC indices, and SRIC and LRIC indices point out countries with considerable outflows of migrants to countries highly involved in international migration and the most interconnected countries.

*Aleskerov Fuad*, National Research University Higher School of Economics (HSE), International Laboratory of Decision Choice and Analysis, V.A. Trapeznikov Institute of Control Sciences of Russian Academy of Sciences (ICS RAS), Moscow; alesk@hse.ru

*Meshcheryakova Natalia*, National Research University Higher School of Economics (HSE), International Laboratory of Decision Choice and Analysis, V.A. Trapeznikov Institute of Control Sciences of Russian Academy of Sciences (ICS RAS), Moscow; natamesc@gmail.com

*Rezyapova Anna*, National Research University Higher School of Economics (HSE), International Laboratory of Decision Choice and Analysis, Moscow; annrezyapova@gmail.com

*Shvydun Sergey*, National Research University Higher School of Economics (HSE), International Laboratory of Decision Choice and Analysis, V.A. Trapeznikov Institute of Control Sciences of Russian Academy of Sciences (ICS RAS), Moscow; shvydun@hse.ru

# 1. Introduction

The role of international migration becomes more important in the modern interconnected world. Migration shapes the world population and influences the society considerably. According to the last report of the General Assembly of the United Nations, number of international migrants worldwide grows faster than the world population: "In 2015, the number of international migrants and refugees reached 244 million, an increase of 71 million, or 41 per cent, from 2000" [28]. The large movements of people will continue or increase due to violent conflicts, income inequality, poverty and climate change: "The world's population is projected to continue to grow for the foreseeable future and is expected to reach 9.7 billion by 2050. If the proportion of international migrants as part of the total population remains constant, the global migrant population will reach 321 million by 2050" [28]. Therefore, international migration is the issue of high importance, and new theories and policies are needed to be developed in order to contribute to the development of both home and host countries.

The international migration theory has a long history starting from Adam Smith [23] in the eighteenth century. Since that time a considerable amount of works has been published in order to explain the causes of migration flows and the consequences of them.

Another approach of studying the process of international migration is a network analysis, in which all countries involved in the international migration are presented as a graph, where nodes are countries and edges correspond to migration flows between them. This approach allows to consider the flows between any two countries integrated into the whole system of countries and shows how the changes in one flow may effect the flows between the other seemingly unrelated countries.

Our work is aimed to detect the countries with highest level of importance in the international migration network. For this purpose we evaluate the classical and new centrality indices. Classical centrality indices are the fundamental attribute of the network analysis and are essential for the representation of major migration flows occurred within the network in a given period. Nevertheless, there is a necessity to consider indirect connections between the countries and node attributes. We use the Indices of Short-Range and Long-Range

Interactions Centralities that take into account the node attributes – population of the destination country as well as indirect connections between the countries in the network.

The paper is organized as follows. Section 2 provides a survey of the literature of international migration analysis. Section 3 gives information on the dataset, its main features and criteria for distinguishing international migrants from tourists and other people, crossing international borders. In Section 4 we describe our methodology and give an interpretation of indices used for the analysis of international migration. In Section 5 we provide the main results of our research. Section 6 concludes.

**Acknowledgement**

The paper was prepared within the framework of the Basic Research Program at the National Research University Higher School of Economics (HSE) and supported within the framework of a subsidy by the Russian Academic Excellence Project '5–100'. The work was conducted by the International Laboratory of Decision Choice and Analysis (DeCAn Lab) of the National Research University Higher School of Economics.

## 2. Literature review

Migration is one of the fundamental processes in the society, therefore it was studied by researchers from the various fields of science: economics, statistics, demography, sociology and mathematics. The literature, which influenced our work, can be divided into two groups: first – theories studying migration on country or country-to-country level, and second – the application of social network analysis to international migration flows.

Migration and its fundamental aspects were studied since early times. Remarkably, one of the first scientists who began to study the process of migration was Adam Smith. The main cause of migration flows between the rural and urban areas according to the hypothesis of A. Smith is that in these areas the wage difference is greater than difference in goods' prices. Additionally, A. Smith compared migration flows to the trade flows and came to a conclusion that trade flows are more intense than migration flows, because migration has more barriers: "man is of all sorts of luggage the most difficult to be transported" [23].

The theory developed after Adam Smith was presented in the "Laws of migration" by E. Ravenstein [21] based on the British population census, migration statistics and vital statistics. All empirical observations E. Ravenstein formulated in 11 Laws of migration, which explain the migration flows. The most relevant for our research statements are 1) the majority of migrants move on short distances, 2) huge migration wave generate the compensating counter wave of migrants, 3) cities with fast growing population are inhabited with migrants from the close rural areas, and the migrants from more distant areas populate the shortage generated in rural areas.

The gravity model of migration plays a significant role in studying migration flows. The model is based on Newton's law of gravitation between two bodies that was applied to the study of migration processes between two countries. In [31] the theory was proposed stating that the level of migration between two territories (Y) is positively related to the population of them and inversely related to the distance between them

$$Y = P_1 \times P_2 / D_{12},$$

where $P_1$ – population of country of the origin, $P_2$ – population of the destination country, and $D_{12}$ – the distance between origin and destination countries.

The intuition behind this hypothesis is rather simple. The inverse relation to distance is explained by the fact that with increasing distance the cost of journey for migrant rises, which negatively affect the level of migration flow. The positive relation to the population of country of origin has the following interpretation. There is a share of population intended to migrate and with the growth of country's population this amount of people increases correspondingly. Finally, if the population of destination country increases, the number of potential employment places and opportunities for migrants enlarges, which make this country attractive for immigrants.

The gravity model became widespread after its application to international trade flows [24]. In this case the gross domestic product (GDP) of two countries is taken into account instead of population. These models are applied in contemporary works explaining international migration flows, for instance, in [25] that will be described later.

Several works explore the phenomenon of migration from the prospect of motives to migrate. The push-pull factors theory has a great importance for the analysis of causes of migration flows [17]. According to that work, there are 4 groups of factors that influence the level of migration between two coun-tries: pull and push factors which characterize both the country of origin and destination, personal factors and intervening obstacles. The examples of the pull factors of the destination country are high wage, high demand for the labor force, considerable amount of social allowance, stable political situation and favorable climate conditions. On the contrary, low wage, unemployment and the conflicts in the country of origin are the push factors for migrants. Personal factors can be different and are defined for each migrant individually. The intervening obstacles can be the huge distance between two countries or strict migration laws.

Migration from the prospect of the economic theory and human capital approach was studied in [22]. It was the first application of the idea of human capital to the field of migration studies [5]. The key logic behind this theory is that a migrant chooses a location that maximizes the net return on migrant's human capital. In this case, the problem lies to the maximization of the individual's profit π from migration from region $A$ to $B$ in each period. It is assumed that there are wage differences between the regions and that a migrant will retire within $T$ periods. Hence, in discrete time the profit from migration from $A$ to $B$ is

$$\pi = \sum_{t=1}^{T} \frac{(W_t^B - W_t^A)}{(1+i)^t} - \sum_{t=1}^{T} \frac{\left(CL_t^B - CL_t^A\right)}{\left(1+i\right)^t} - C\left(D, X\right),$$

and in continuous time

$$\pi = \int_{t=0}^{T} [W_t^B - W_t^A - CL_t^B + CL_t^A] e^{rt} dt - C(D, X),$$

where $W_t^B$ and $W_t^A$ – wages in destination and origin countries accordingly, $CL_t^B$ and $CL_t^A$ – costs of living in region $B$ and $A$, $i$ and $r$ – the interest rates, $C$ – costs of migration from $A$ to $B$, which depend on the distance between regions ($D$) and any other factors influencing on costs ($X$). In both discrete and continuous time models an individual is willing to migrate from $A$ to $B$ only if his/her profit will be positive, i.e., $\pi > 0$.

The human capital model of migration became fundamental for many modern models aimed to study migration from different aspects. The later works take into account more factors influencing the migration: the influence of kinship and migrant network [30], introducing a family as a decision-making unit [19], studying migration decision in a life-cycle context [20], and imposing remittances as another factor influencing migration [9]. For more detailed review of migration theories see [5].

The theories reviewed above apply different levels of analysis of human migration: the macro-level (migration between countries and regions) and micro-level (individual). However, they have a common attribute: migration is a bilateral process and migration flows between any two countries are studied independently from the flows between other countries.

The process of migration is complex and the level of migration between any two countries depends not only on factors related to these two countries, but also on migration flows between other countries. In the network analysis countries are not isolated elements, all of them are interconnected through migration flows. The migration process is modeled as a weighted-directed graph, where nodes are countries and edges – migration stocks or flows between them.

The application of the network approach to the international migration was presented in [11]. The data was taken from the World Bank international migration database [34] for each decade of the period from 1960 to 2000. The database contained information about stock of migrant population in 226 countries, i.e. people living in the country other than a country of their origin in a given point of time.

In that study the International Migration Network (IMN) is constructed as weighted-directed graph, where nodes are countries and edges correspond to stock of migrants. Interesting findings were obtained by analyzing binary and weighted characteristics of the network, clustering based on network structure and gravity modeling.

Weighted-network statistics had power-law distribution, meaning that migrant stock was increasing over time. Additionally, the number of connections also had increased over the period; countries became more interconnected through migration flows, which corresponds to the trends in international migration [27].

IMN was characterized as a network with high clustering and disassortativity. This result is rather simply interpreted empirically. High clustering relates to connections between countries over time. The following clusters of countries were formed: Asian and Sub-Saharan African, former Soviet Union, European and American. Disassortativity in IMN means that countries with low migrant stock are likely to be connected with countries with huge migrant population, i.e. there are established countries of migrant origin and destination.

The results of ordinary least squares regression and gravity model outline geographical, political and socio-economic factors as more significant than local network properties for the structure of IMN.

International migration was studied in [10] by constructing the global human migration network. The data on migrant stock for 226 countries were used as in [11]. The data were available for the period from 1960 to 2000 for each decade. Interesting characteristics of the global human migration network, community analysis and the development of the network over the period were introduced.

In [10] properties of the migration network presented as a weighted-directed graph were analyzed and the following results were obtained. The largest connections were found within Europe, between Middle East and India, within former Soviet Union countries, from Western Europe, Canada, Eastern Asia and Mexico to the United States of America (USA). The results do not perfectly correspond to the growing issue of "South-North" migration, as were remarked in [10]. Communities of countries with intense connections within them and modest inter-community connections were formed and appeared to be very similar to the communities identified by [11]. The global human migration network turned out to increase in interconnection and transitivity and decrease in average path's length over the period. These results are highly related with the processes of globalization and escalation of human mobility over the past time.

Overall, in [10, 11] the fundamental network analysis of the international migration was proposed with the results having meaningful empirical evidence. The analysis in both papers is based on the migrant stock statistics, which is an accumulative pattern that represents total number of migrants living in a given country in certain period. However, there is another statistics of international migration, which represents the flow of international migrants arriving to a given country or leaving it each year. In our work we use the database on international migration flow provided by United Nations (UN) [32, 33].

One of the most recent and relevant papers which studies migration flows from the network prospective is [25]. The research is focused on the network analysis of international migration flows between countries of the Organization for Economic Co-operation and Development (OECD) (32 countries). The analysis can be divided into the following steps: estimation of the network attributes, community detection in international migration network, and, finally, application of the generalized gravity model to international migration flows using panel data regressions and multivariate regression quadratic assignment procedures.

As network attributes several centrality indices (degree, weighted degree, normalized weighted degree) were estimated for one year period (2000) and some interesting features of the international migration network were obtained.

Degree centrality characterizes the number of countries connected with the given country through migration flows. The USA, Canada and some European countries (Austria, Finland, Spain, Sweden) have the highest in-degree centrality. In other words migration flows to these countries are originated in the highest number of different countries. The USA, The United Kingdom (the UK) and Germany had the highest out-degree centrality, i.e. the number of countries-destinations for migrants from these countries was the highest. The USA, Canada and Germany were ranked as top-3 by degree centrality and had the in-flow and out-flow of migrants to the largest number of countries.

The next group of centrality indices evaluated in that work are weighted degree centralities, which consider the number of migrants in inter-country migration flows. Weighted in-degree centrality is the number of immigrants and weighted out-degree is the number of emigrants for each country. In addition, the difference between in-degree and out-degree was calculated, which stands for the net migration flow. The USA, Germany and the UK had the highest number of migrant in-flow, Mexico, Poland and the UK were top 3 countries of migrant out-flow, and Germany, the USA and Switzerland had the highest net migrant flow.

Another step in that paper was the normalization of weighted degree centralities by the population of destination country. The normalization is important in the context of understanding the influence of immigration flow on the country of destination: the flows of 5000 people for countries with the population of around 0.5 million people (e.g., Luxembourg) and 300 million people (e.g., the USA) produce completely different effect. For example, Luxembourg, Switzerland and Germany are top 3 countries in ranking by normalized weighted in-degree, which is different from the top 3 countries by weighted in-degree centrality (Germany, the USA and the UK). The population of destination country is an essential network attribute used in our research.

Earlier theories reviewed above provide the fundamental understanding of factors influencing the migration, which is essential to analyze the temporary processes emerging in the society. Recent studies show that internation-al migration became more complex process, where connections between countries are strengthening and new connections are developing. Consequences of changes in human mobility in certain directions to the entire network of countries can be dramatic. Therefore, it is important to study international migration from the network perspective and find the countries with considerable influence on the whole network through migration flows.

## 3. The data

Data on international migration is usually presented in two fundamental statistical categories: stock of migrants and migration flows. Migration flow is defined as a number of persons arriving to country or leaving it in a given time period. Migrant stock corresponds to the total number of people living in a country other than the country of origin in a certain moment. The key difference between these two categories is that the stock of migrants is an accumulative pattern, and the flow data represents the fact of immigration or emigration to or from a given country.

We use the data on migrant flow for an analysis of the international migration. The high frequency flow statistics is extremely difficult to find. Additionally, it becomes even more challenging when the research is focused not on the analysis of the migration within the certain geographical region or the association of countries, but on the international migration worldwide. The data provided by the United Nations Organization [32, 33] is rather helpful, when the purpose is to maximize the number of included countries. Therefore, the UN international migration flow statistics was used. However, international migration flow data usually lacks completeness and is collected by the national statistical agencies for various political purposes. These factors lead to difficulties in possibilities of making cross-country comparisons and inconsistency in data.

Next, we provide the description of the database and the steps accomplished to resolve the problem of inconsistency in data.

### *3.1. Data Description*

Two datasets, both collected by United Nations Population Division: 2009 Revision and 2015 Revision [32, 33] were used for the construction of inter-national migration network. These datasets contain time series dyadic data on migration flows from selected countries.

The 2008 Revision included data on international migration flows from 29 countries for the period from 1970 to 2008. The 2015 Revision was characterized by the increase in the number of countries to 45 and different period (from 1980 to 2013). The list of countries that provided statistics for each database is presented in Tables 15 and 16 in the Appendix. Migration flows for countries not included in the list were accumulated by the statistics of the countries presented in each database.

To distinguish international migrants from other categories of movers, countries apply different time criterion – the minimal period of staying abroad. By this criterion countries are divided into the following groups: establishment of permanent residence (abroad), expected stay (abroad) of at least one year, six months, three months, other time criterion or they do not specify it.

The data was collected through different sources: population registers, border statistics, the number of residents permits issued, statistical forms that persons fill when they change place of residence and household survey.

There are three ways to define country of migrants' origin or destination by

1) residence; 2) citizenship; 3) place of birth.

The distribution of countries from the two databases by these criteria is presented in Table 1.

*Table 1.* Distribution of countries by country of origin criterion

| **Number of Countries by** | **Datasets** | | | |
|---|---|---|---|---|
| | **v2008** | | **v2015** | |
| | Inflows | Outflows | Inflows | Outflows |
| Citizenship | 7 | 7 | 36 | 37 |
| Residence | 21 | 21 | 43 | 44 |
| Place of Birth | 1 | – | 1 | – |

Most countries in both 2008 and 2015 Revisions define the country of origin as the country of previous residence. However, statistics differs in both datasets for inflows and outflows.

For 2008 Revision 21 countries (Australia, Austria, Canada, Croatia, Czech Republic, Denmark, Estonia, Finland, Germany, Iceland, Israel, Italy, Latvia, Lithuania, New Zealand, Norway, Poland, Slovakia, Spain, Sweden, the UK) apply residence criterion to define the country of origin or destination. In seven countries (Belgium, France, Hungary, Luxembourg, Netherlands, Slovenia, Switzerland) the country of citizenship was used to classify migrants, and only in the USA the place of birth was used to define the origin of migrants.

There are considerable differences in distribution of countries by this criterion in the 2015 Revision compared to 2008: for 43 out of 45 countries there are data on migration flows based on residence. This list lacks only the USA and Canada, where place of birth and citizenship criteria were used correspondingly.

Additionally, as countries apply different criteria to determine main concepts concerning international migration, there were some cases of inconsistency in observations. The steps proposed to make data more comparable are presented below.

### *3.2. Data aggregation*

There were three key issues in aggregation of the databases: the choice of the most relevant criteria on the country of origin, inconsistency in data on the certain migration flows, and the cases of flows with the same country of origin and destination.

The preference was given to statistics on residence, when data for both residence and citizenship were available. The reason is that, as we can see from Table 1, more countries apply this criterion in the 2015 version. Additionally, this principle more accurately reflects the definition of the international migrant by the United Nations Organization: person who changes his or her country of usual residence. Country of citizenship is not mandatory the country of usual residence and country, where migrant lived before (previous residence), that is why data on residence is more representative in terms of migration flows. Overall, about 80% of migrant flows are characterized by the previous residence of migrants, 16% – by their citizenship and only 4% – by their place of birth.

The preference for the 2015 Revision was given as well, when there was the data from both datasets. An exception is the case, when there are data based on residence in 2008 version and no data on residence in 2015 version.

Another important issue is the inconsistency in the same migration flows. Overall in 5% of observations data was inconsistent: for the same migration flow data from different countries was not the same (8 672 out of 173 435 observations). In most of these cases the difference was not significant, therefore the mean value was taken. However, there were 21 observations, where simultaneously the minimum value was less than 10 and the ratio between maximal and minimal provided value was more than 1 000. All these cases were studied individually, and at the end were explained by incorrect statistics in data of country with minimal value. Thus, only maximum values were taken into account. For example, for 5 observations the country of destination or origin is one of the former Soviet Union countries. After Soviet Union disintegration migration statistics in these countries was not of the high quality compared to the data of other countries. A list of the inconsistent observations is provided in Table 17 in the Appendix.

Another feature of the dataset was the presence of flows, which have the same country of migrant origin and destination (loops in terms of networks). Total number of loops in aggregated data was 743. The documentation of the 2008 Revision [32] provides the following explanation for some of them. For Sweden and Spain: the criterion for the country of origin was citizenship, thus, these migrants were returned citizens. These observations are not important for our study, because they do not contain the information about previous migrant's location. For Australia loops in dataset were explained as migration flows between Australia and its external island territories or internal migration. This data is not applicable for international migration flows. Other countries did not provide the information about such cases, thus we assume that the explanation is similar to one of the given above. Therefore, we can conclude that the cases of the same origin and destination countries can be excluded from observations, as they do not have any meaningful interpretation.

To conclude, aggregation of two Revisions [32, 33] was made, the problem of inconsistency in observations was resolved, loops were eliminated and as a result, the annual data on international migration flows from 1970 to 2013 for 215 countries was obtained.

## 4. Centrality indices

International migration patterns are usually analyzed by simple measures as the number of migrant inflows and outflows, net and gross migration flows. These measures are basic and can be useful for a certain country concerning its migration policy. However, global migration forms the network of coun-tries and all of them are interconnected through migration flows. Therefore, in our analysis international migration is modeled as a graph, where nodes are countries and edges show migration flows.

We study the properties of international migration flows from the network prospective, evaluating the centrality indices. The aim of this methodology is to provide a ranking of the countries based on their importance for the migration process.

First, we apply classical centrality indices to international migration. Second, we propose to use new centrality indices with certain distinctive features in comparison with classical centrality indices.

### *4.1. Classical centrality indices*

In our work the following centrality measures are evaluated: degree and weighted degree centrality, closeness, eigenvector and PageRank.

The degree centrality is the number of nodes each node is connected with [13]. For directed graph the degree centrality has three forms: the degree, in-degree and out-degree centrality. The in-degree centrality represents the number of in-coming ties each node has, and out-degree is the number of out-going ties for each node.

In terms of migration, edge in unweighted graph characterizes the presence of migration flow between any two countries. The in-degree centrality for country *A* is the number of countries, which are connected with country *A* through migration in-flows to country *A*. In other words, it is the number of countries, which migrants came to country *A* from. For out-degree centrality the number of countries is evaluated, which are connected with country *A* through migrant out-flows from *A*, i.e. the number of countries which are the destinations of migrants from *A*. The degree centrality of country *A* can show how many different countries are connected with it through migration flows.

The following centrality indices were estimated for the weighted network: weighted in-degree, weighted out-degree, weighted degree difference (=weighted in-degree – weighted out-degree) and weighted degree [13]. The *weighted in-degree* (WInDeg) centrality represents the number of in-coming ties for each node with weights on them, i.e. the immigrant flow to the country. *Weighted out-degree* (WOutDeg) is the number of out-going links for each node and accordingly relates to the number of emigrants. The *weighted degree difference* (WDegDiff) is the difference between migrant in-flow and out-flow which is the net migration flow. The *weighted degree* is the sum of weighted in-degree and weighted out-degree centralities for each country, i.e. the total number of emigrants and immigrants (gross migration). These centrality indices can give us the basic information about the international migration process: the level of migrant in-flows and out-flows, net and gross migration flows.

The *closeness* (Clos) [4] centrality shows how close a node is located to the other nodes in the network. In addition, this measure has the following characteristics. Firstly, it accounts only for short paths between nodes. Secondly, these centralities have very close values and are sensitive to the changes in network structure: minor changes in the structure of network can lead to significant differences in raking by this measure. In our work the closeness centrality is estimated for the undirected graph with maximization of the weights on paths and is related to the level of closeness of particular country to intense migration flows. Note that it does not imply that the country itself should have huge migration in-flows or out-flows. This measure can provide the information about potential migration flow to particular country by estimation the distance between the country and countries with huge migration flows in the network. Countries with low closeness centrality value are not necessarily involved in the process of international migration since they usually have low migration flows.

*Eigenvector* (Eigenvec) [7] is the generalized degree centrality, which accounts for degrees of node neighbors. Eigenvector centrality and its analogue *PageRank* [8] centrality measure are based on the idea that a particular node has a high importance if its adjacent nodes have a high importance. In international migration network these indices highlight the countries – "centers of international immigration", and the countries, which are directly linked with them through migration flows.

### 4.2. Short-Range Interaction and Long-Range Interaction Centrality indices

Short-Range (SRIC) and Long-Range Interactions Centrality (LRIC) indices have the following distinct features. They account for the indirect interactions between countries and population of the destination country.

The indirect influence of country A to country B through migration flows is important to consider in the network for the following two reasons. First, migration between any two countries may occur not directly, i.e. there can be a migration route. In this case the understanding of country with highest indirect influence, i.e. the initial country generating the migration flow is meaningful to highlight the most powerful countries in the global migration network. Second, as all countries in international migration network are interconnected, the flow between any two countries can lead to emergence of new flows between any other countries. In this case flows of migrants do not necessarily consist of the same people, as we do not know migrants' characteristics (nationality, gender and other). Both cases are possible in the analysis of indirect influence of countries in the network. However, classic centrality indices do not consider the indirect interactions.

*Short-range interaction centrality index (SRIC)* is based on the power index proposed in [1] and applied for networks in [2]. The key difference of this index from classic centrality indices is that it takes into account node attributes (the population of country in our case), and indirect influence between them.

We evaluate the direct influence of one country to another one through imposing the quota, which represents the population of the destination country. We suggest that 0.1% of population of destination country is the critical level of migrant inflow. If the migration flow from country $A$ to country $B$ does not reach 0.1% of population of country $B$, then country $A$ does not directly influence country $B$ through migration flows.

The critical group of countries is interpreted as a group whose total number of migrants is critical in terms of quota for the population of destination country, i.e. the group is critical if the total number of its members' immigrants is greater than or equal to a predefined quota. A country is pivotal in the critical group, if without this country group is no longer critical. The intensity of connections $f\left(i, w_a\right)$ is estimated by the following formula

$$f\left(b, w_a\right) = \frac{p_{ba} + p'_{ba}}{|w_a|},$$

where $w_a$ is a critical group of countries with respect to a country $A$ (country of origin), in which a country $B$ (destination country) is pivotal, $p_{ba}$ is the total number of migrants came from country $A$ directly to $B$, $p'_{ba}$ is the total number of migrants came from country $A$ to $B$ indirectly – via any other country. Below a simple example is presented of the different indirect paths from country $A$ to country $B_3$ for the Short-Range Interaction Centrality Index.

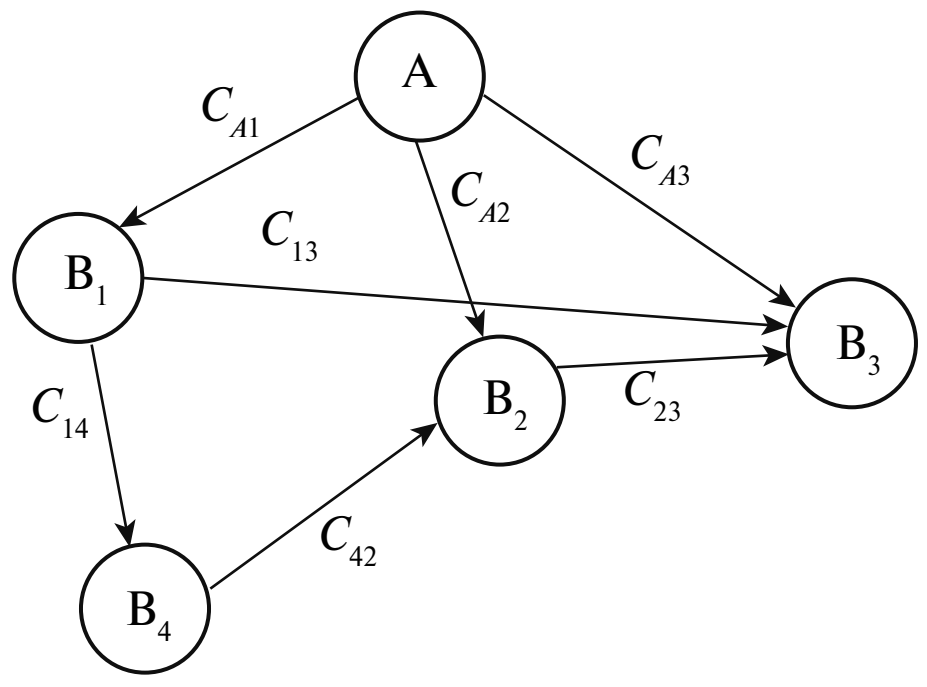

**Fig. 1.** Direct and indirect influence between elements

As we can see from the graph, there are three different ways to reach $B_3$ from $A$: 1) $A$-$B_1$-$B_3$, 2) $A$-$B_2$-$B_3$ and 3) $A$-$B_1$-$B_4$-$B_2$-$B_3$. SRIC accounts only for the first order connections as in the cases 1) and 2). However, migrants from $A$ can move to $B_3$ using longer route and in this case we need to re-evaluate the estimation of index to consider s-Long-Range routes (in this case 3-long range routes). Thus, we use another index that takes into account these features.

*Long-Range Interaction Centrality Index (LRIC)* was proposed in [3]. LRIC is estimated as follows.

First, the matrix of bilateral migration flows is constructed $A = [a_{ij}]$, where $a_{ij}$ is the migration flow from country $i$ to country $j$. Then we construct a matrix $C = [c_{ij}]$ with respect to the matrix $A$ and predefined quota as

$$c_{ij} = \begin{cases} \dfrac{a_{ij}}{\min\limits_{\Omega(i)\subseteq N_i | j\in\Omega_p(i)} \sum_{l\in\Omega(i)} a_{il}}, if\ j \in \Omega_p(i) \subseteq N_i, \\ 0, j \notin \Omega_p(i) \subseteq N_i, \end{cases}$$

where $\Omega(i)$ is a critical group of direct neighbors for the element $i$, $\Omega(i) \subseteq N_i$, and $\Omega_p(i)$ is a critical group for the element $i$, $\Omega_p(i) \subseteq \Omega(i)$. A group of neighbors of the node $i$ $\Omega(i) \subseteq N_i$ is critical if $\sum_{l\in\Omega(i)} a_{il} > q_i$.

Obviously, the construction of matrix $C$ is highly related to [2] because it requires to consider separately each element of the system as a country of des-tination while other participants of the system are assumed as countries of migrants origin.

The interpretation of matrix *C* is rather simple. If $c_{ij} = 1$, then the country of migrants origin *j* has a maximum influence to the country of migrants destination *i*. On the contrary, if $c_{ij} = 0$ then the country of origin *j* does not directly influence the country of destination *i*. Finally, the value $0 < c_{ij} < 1$ indicates the level of impact of the origin country *j* on the destination country *i*.

Thus, we evaluate the direct influence of the first level of each element in the system. To define the indirect influence between two elements we should consider all possible paths between them. A path from *i* to *j* is an ordered sequence of steps starting at *i* and ending at *j*, such that the second element in each step coincides with the first element of the next step. In other words, it is an ordered sequence of elements $i, j_1, \ldots, j_k, j$, such that $i\rho j_1, j_1\rho j_2, \ldots, j_{k-1}\rho j_k, j_k\rho j$, where $j_1 \rho j_2 \Leftrightarrow c_{j_1 j_2} > 0$. The number of steps in a path is called the path's length. Additionally, we can limit the path's length by some parameter *s*.

We consider only paths with no cycles, i.e. there are no elements that occur in the path at least twice. Denote by $P^{ij} = \{P_1^{ij}, P_2^{ij}, \ldots, P_m^{ij}\}$ a set of unique paths from *i* to *j*, where *m* is the total number of paths and denote by $n(k) = \left|P_k^{ij}\right|$, where $k = \overline{1,m}$, a length of the *k*-th path. Then we can define the indirect influence $f\left(P_k^{ij}\right)$ between elements *i* and *j* via the *k*-th path $P_k^{ij}$ as

$$f\left(P_k^{ij}\right) = c_{ij(1,k)} \cdot c_{j(1,k)j(2,k)} \cdot \ldots \cdot c_{j(n(k),k)j}, \tag{1}$$

or

$$f\left(P_k^{ij}\right) = \min(c_{ij(1,k)}, c_{j(1,k)j(2,k)}, \ldots, c_{j(n(k),k)j}), \tag{2}$$

where $j(l,k)$, $l = \overline{1,n(k)}$ is an *l*-th element which occurs on *k*-th $\rho$-path from *i* to *j*.

The interpretation of formulae (1) and (2) is the following. According to the formula (1) the total influence of the element *j* to the element *i* via the *k*-th $\rho$-path $P_k^{ij}$ is calculated as the aggregate value of direct influences between elements which are on the *k*-th path between *i* and *j* while the formula (2) defines the total influence as the minimum direct influence between any elements from the *k*-th path.

Since there can be many paths between two elements of the system, there is a problem of aggregating the influence of different paths. To estimate this aggregated indirect influence several methods are proposed.

The aggregated results will form a new matrix $C^*(s) = [c^*_{ij}(s)]$.

*1. The indirect influence: sum of paths influence*

$$c^*_{ij}\left(s\right) = \min(1, \sum_{k:\, n(k)\le s} f\left(P^{ij}_k\right)). \tag{3}$$

*2. The indirect influence: maximal path influence*

$$c^*_{ij}(s) = \max_{k: n(k)\le s} \max_{k:\, n(k)\le s} f\left(P^{ij}_k\right). \tag{4}$$

Thus, the sum of paths influences gives the most pessimistic evaluation of the indirect influence where we take into account all possible channels of migration from a particular origin country to the country of destination.

We can define the indirect influence between elements *i* and *j* via all possible paths between them. The paths influences can be evaluated by formulae (1)–(2) and aggregated into a single value by formulae (3)–(4). Four combinations are possible for matrix $C^*(s)$ construction (see Table 2). In our opinion, all possible combinations of the formulae have a sense except the combination of formulae (2) and (3).

*Table 2.* Possible combinations of methods for indirect influence

| **Influence/Aggregation** | | **Paths aggregation** | |
|---|---|---|---|
| | | **Sum of paths influence** | **Maximal path influence** |
| **Path influence** | **Multiplication of direct influence** | SumPaths | MaxPath |
| | **Minimal direct influence** | – | MinMax |

The aggregation of matrix $C^*(s)$ into a single vector showing the total influence of each element of the system can be done with respect to the weights (importance) of each element as it is done in [2].

To sum up, the classic centrality indices and indices of Short and Long-Range Interactions Centralities are applied to characterize the countries in migration network. The distinctive feature of the latter is the consideration of the population of destination country and indirect migration routes between countries.

# 5. The results

The centrality indices are evaluated for each year of the period 1970–2013. The results are presented in the following form. In the Subsection 5.1 we compare the ranking of countries by the classical centrality, SRIC and LRIC indices. In the Subsection 5.2 the description of dynamics of centrality indices is provided.

## *5.1. Ranking by classic and Short-Range Interactions and Long-Range Interaction Centralities*

The analysis for each decade is presented in the following form. First, the overall picture of the international migration in the corresponding decade is observed by overview of the major migration corridors. Second, the results of evaluation of classic centralities and SRIC, LRIC indices are presented. Finally, the comparison of results is made by performing the correlation analysis.

**1970–1979**

The major migration corridors for 1970s occurred between Turkey and Germany (in both directions), Yugoslavia and Germany (in both directions), within the European countries, from Mexico to the USA and from the UK to Australia. The migration ties between developed European countries and developing countries during this period is explained by the labor migration program [29]. This program influenced the migrant inflow from south European countries (Italy, Spain and Greece) and developing countries outside European region (Turkey). The situation changed after the oil crisis in 1973. The guest labor migration program was over and it caused the emigration of people, which already were unemployed. Additionally, in this decade the migration flow from Mexico to the USA begins to exceed 50 000 of migrants since 1972. On the contrary, migration from the UK to Australia drops and after 1974 is no longer presented in the list of corridors over 50 thousand of migrants.

Now we present the results of evaluation of centrality indices. As was mentioned above, major migration corridors did not change considerably, hence centrality indices did not differ a lot during these years. In order to represent the international migration flows in the 1970s from the perspective of centrality indices the 1972 results were chosen.

To begin with, migration flows over 50 thousands for 1972 are presented in Table 3. This list complies the major migration corridors occurred in the 1970-1979. Consequently, let us provide the ranking of countries by the centrality indices (Table 4).

*Table 3.* Migration flows over 50 000 in 1972

| Origin | Destination | Migration flow |
|---|---|---|
| Turkey | Germany | 161 430 |
| Germany | Italy | 122 888 |
| Germany | Turkey | 111 401 |
| Germany | Yugoslavia (former) | 102 588 |
| Italy | Germany | 88 062 |
| Yugoslavia (former) | Germany | 72 835 |
| Mexico | USA | 71 586 |
| UK | Australia | 63 800 |

Germany was involved in the largest migration flows as migrant destination and origin country. Therefore, it has highest weighted in-degree centrality (migrant inflow), weighted out-degree (migrant outflow), weighted degree (gross migration flow), correspondingly. Weighted in-degree centrality also highlighted the immigration countries: Italy, Yugoslavia (former) and the UK.

Weighted out-degree centrality results correspond to the countries-suppliers of the labor force – Turkey and Italy. The UK was in the top of countries by this centrality because of the flow to Australia.

The highest weighted degree centrality or the gross migration flow had most involved into the process of international migration countries (Germany, the USA, Italy, Turkey and Yugoslavia (former)). The USA are constantly ranked the first by weighted degree difference (net migration flow). This fact is explained not only by the attractiveness of this country for migrants, but that the USA do not provide the emigration statistics, hence net migration flow does not contain this component.

Closeness centrality ranks the countries based on the presence of connections with main migrants' origin or destination countries. The new country introduced by this centrality is Sweden, because there were emigration from Sweden to both the USA and Germany.

*Table 4.* Rankings by centrality indices for 1972

| Country | WInDeg | WOutDeg | WDeg | WDegDiff | Clos | PageRank | EigenVec | SRIC | LRIC (SUM) | LRIC (MAX) | LRIC (MAXMIN) |
|---|---|---|---|---|---|---|---|---|---|---|---|
| Germany | 1 | 1 | 1 | 2 | 2 | 1 | 1 | 1 | 5 | 4 | 4 |
| USA | 2 | 6 | 2 | 1 | 1 | 2 | 6 | 2 | 4 | 6 | 8 |
| Italy | 3 | 3 | 3 | 212 | 5 | 4 | 2 | 4 | 3 | 3 | 3 |
| Yugoslavia (former) | 4 | 5 | 5 | 211 | 6 | 6 | 3 | 6 | 1 | 1 | 1 |
| UK | 5 | 4 | 6 | 213 | 13 | 3 | 9 | 7 | 9 | 11 | 12 |
| Canada | 6 | 18 | 7 | 3 | 7 | 5 | 12 | 11 | 14 | 17 | 15 |
| Turkey | 7 | 2 | 4 | 215 | 3 | 7 | 4 | 3 | 2 | 2 | 2 |
| Australia | 8 | 12 | 10 | 4 | 25 | 8 | 16 | 5 | 12 | 12 | 11 |
| Greece | 9 | 7 | 8 | 205 | 10 | 9 | 5 | 16 | 6 | 5 | 5 |
| Spain | 10 | 9 | 9 | 198 | 11 | 11 | 7 | 12 | 7 | 7 | 6 |
| Netherlands | 11 | 15 | 11 | 7 | 9 | 12 | 10 | 15 | 13 | 13 | 13 |
| Belgium | 12 | 16 | 12 | 9 | 16 | 10 | 13 | 19 | 22 | 21 | 16 |
| Sweden | 13 | 13 | 13 | 199 | 4 | 13 | 19 | 8 | 17 | 20 | 27 |
| Austria | 14 | 11 | 14 | 202 | 14 | 14 | 8 | 18 | 8 | 8 | 7 |
| France | 15 | 14 | 16 | 204 | 17 | 16 | 11 | 10 | 10 | 9 | 9 |
| South Africa | 16 | 27 | 20 | 5 | 31 | 15 | 18 | 13 | 20 | 19 | 18 |
| Finland | 17 | 24 | 18 | 8 | 24 | 18 | 23 | 9 | 18 | 18 | 25 |
| New Zealand | 18 | 31 | 23 | 6 | 40 | 17 | 32 | 14 | 25 | 24 | 23 |
| Portugal | 20 | 10 | 17 | 210 | 15 | 20 | 15 | 17 | 11 | 10 | 10 |
| Norway | 23 | 47 | 36 | 10 | 49 | 23 | 33 | 41 | 41 | 41 | 41 |
| Mexico | 42 | 8 | 15 | 214 | 8 | 42 | 45 | 41 | 41 | 41 | 41 |

PageRank and Eigenvector centralities account for attractive migrants' destination countries (Germany, the USA, Yugoslavia (former) and Italy) and, in addition, countries connected with them (the UK, Canada and Turkey).

Overall, ranking by classical centrality indices shows the countries directly involved in the process of international migration: top countries of migrants' destination, origin and their direct neighbors in the network.

Consideration of the indirect interactions can help to outline a new list of countries with high influence in international migration network.

SRIC ranking of countries is highly related with ranking by weighted in-degree centrality. However, Turkey is presented among top three countries, and Finland appears in top ten. Let us explain this results. Turkey has direct connections with Germany through highest migration inflow and outflow. Finland also has migrant inflow and outflow to Germany, nonetheless they are not massive (2 862 and 3 663, correspondingly). They influence Finland, because population of Finland was not very large (4 639 657) in comparison with other countries.

Each LRIC index highlights Yugoslavia (former) and Turkey, as these countries are interconnected with the countries – centers of migrant attraction (Germany, the USA, Italy), which have lower position in ranking. Interestingly, Greece is outlined in top six countries. Greece had both immigrants from Germany (from Germany to Greece 48 538) and sent migrants to Germany (51 509), the USA (11 021) and Canada (4 016). Additionally, population of Greece was 8 888 628 in 1972. Spain and Austria also had higher ranking by LRIC indices. As in the previous cases consideration of indirect interactions of these countries in the network and their population made them rise in ranking. Spain was a labor supply country for Germany, therefore they were connected by both inflows and outflows of migrants in 1972. Austria and Germany had established migration connections because of geographical and cultural proximity.

SRIC and LRIC indices define different from classical centralities rankings of countries. These indices outline not only top migrant origin and destination countries, but also the countries connected with them (Greece, Spain, Austria) and countries, where immigrants have considerable share of the population (Finland).

For comparison of rankings of countries by different centralities, correlation analysis is applied. As the position of country in the ranking is the rank variable Goodman, Kruskal $\gamma$-coefficient [14] was estimated for each year of the period. The results did not vary considerably for each year. Therefore, the estimation results are provided for 1972 (Table 5) as an example.

The ranking by SRIC and LRIC is highly related to eigenvector, PageRank and weighted degree centralities, as was observed after the estimation of Goodman-Kruskal correlation coefficient [14]. Additionally, SRIC and all LRIC indices are highly correlated between each other and weakly with weighted degree difference. However, as it was mentioned in the description of the results above classical centrality indices do not consider countries connected with top migrant destinations and share of the migrants in the population of the country.

*Table 5.* Goodman, Kruskal γ-coefficient for 1972

| | SRIC | LRIC (SUM) | LRIC (MAX) | LRIC (MAXMIN) |
|---|---|---|---|---|
| **WInDeg** | 0.91 | 0.918 | 0.918 | 0.908 |
| **WOutDeg** | 0.874 | 0.881 | 0.88 | 0.877 |
| **WDeg** | 0.889 | 0.89 | 0.89 | 0.885 |
| **WDegDiff** | –0.392 | –0.401 | –0.4 | –0.401 |
| **Clos** | 0.885 | 0.888 | 0.887 | 0.882 |
| **PageRank** | 0.9 | 0.907 | 0.907 | 0.898 |
| **Eigenvec** | 0.897 | 0.92 | 0.921 | 0.91 |
| **SRIC** | 1 | 0.966 | 0.963 | 0.97 |
| **LRIC (SUM)** | | 1 | 0.995 | 0.984 |
| **LRIC (MAX)** | | | 1 | 0.983 |
| **LRIC (MAXMIN)** | | | | 1 |

**1980–1989**

The international migration flows during these decade can be divided into the following groups: 1) from Central America to the USA, 2) from Southeast Asia to the USA, 3) intra-European migration, 4) from Turkey and Yugoslavia (former) to Germany.

Migration flows to the USA from the Central American countries were characterized by the rise of the inflow from Mexico and the development of the new flows from other countries of this region (El Salvador). Also the inflow from the Southeast Asia countries that already occurred in the previous decade became more intense. It represents the immigration of qualified labor force, which receives higher education in their country of origin (the Philippines, Vietnam) and migrate to the USA to provide their families with remittances.

The flows already established in the previous period – from Turkey and Yugoslavia (former) to Germany – are still presented and there is a considerable rise in Poland to Germany migration caused by the economic and political crisis in Poland.

The situation changed in the end of the 1980s: new migration flows were generated by Germany reunification and economic and political crisis in the USSR in 1989. The first caused the migration corridor between German Democratic Republic (former) and German Federal Republic. The USSR crisis in economy and politics was officially declared in 1989 and caused the wave of emigration to Germany.

In Table 6 below the major migration flows for 1989 are provided.

*Table 6.* Migration flows over 50 000 in 1989

| Origin | Destination | Migration flow |
|---|---|---|
| Poland | Germany | 455 075 |
| Mexico | USA | 405 172 |
| German Democratic Republic | Germany | 388 396 |
| Germany | Poland | 145 903 |
| USSR (former) | Germany | 121 378 |
| Turkey | Germany | 86 643 |
| Yugoslavia (former) | Germany | 63 438 |
| El Salvador | USA | 57 878 |
| Philippines | USA | 57 034 |

Considering the overview of the migration flows mentioned above it is interesting to view how these processes were described by the centrality indices. Till the end of the decade the ranking of countries by centralities did not change a lot. At the end of the decade because of the changes in international migration mentioned above the results of centrality indices also evolved dramatically for 1989 (Table 7).

Countries with the highest inflow of migrants – Germany, the USA, the UK and Australia were the leaders in weighted in-degree, weighted degree and weighted degree difference. From the ranking by weighted out-degree centrality the following countries with largest outflow of migrants are presented: Poland, Germany, Mexico, the UK and the USSR (former). Weighted Degree centrality do not outline any new countries. The largest net migration flow had Germany, the USA, Australia and the UK.

Canada, the USA, Netherlands and Denmark have the highest ranking by closeness centrality. The USA are in the top of ranking because of the huge inflows. Other countries had outflows of migrants to the USA or to Germany, both were the international immigration centers.

*Table 7.* Rankings by centrality indices for 1989

| Country | WInDeg | WoutDeg | Wdeg | WdegDiff | Clos | PageRank | EigenVec | SRIC | LRIC (SUM) | LRIC (MAX) | LRIC (MAXMIN) |
|---|---|---|---|---|---|---|---|---|---|---|---|
| Germany | 1 | 2 | 1 | 2 | 10 | 1 | 1 | 2 | 7 | 6 | 6 |
| USA | 2 | 8 | 2 | 1 | 2 | 2 | 3 | 9 | 4 | 7 | 7 |
| UK | 3 | 4 | 4 | 5 | 16 | 3 | 7 | 6 | 12 | 13 | 16 |
| Australia | 4 | 6 | 6 | 4 | 9 | 4 | 11 | 7 | 10 | 16 | 14 |
| Canada | 5 | 27 | 7 | 3 | 1 | 6 | 8 | 35 | 29 | 43 | 35 |
| Poland | 6 | 1 | 3 | 214 | 11 | 5 | 2 | 3 | 5 | 4 | 4 |
| Netherlands | 7 | 23 | 11 | 6 | 3 | 7 | 12 | 29 | 22 | 27 | 36 |
| Italy | 8 | 11 | 9 | 8 | 18 | 9 | 6 | 15 | 8 | 9 | 8 |
| Sweden | 9 | 40 | 15 | 7 | 8 | 10 | 24 | 13 | 44 | 50 | 52 |
| New Zealand | 10 | 14 | 13 | 182 | 6 | 8 | 32 | 12 | 15 | 14 | 13 |
| Turkey | 11 | 7 | 8 | 211 | 20 | 12 | 4 | 5 | 2 | 2 | 2 |
| Yugoslavia (former) | 12 | 10 | 12 | 203 | 29 | 14 | 5 | 16 | 6 | 5 | 5 |
| Denmark | 13 | 24 | 19 | 10 | 4 | 11 | 28 | 18 | 42 | 46 | 48 |
| Spain | 14 | 37 | 22 | 9 | 17 | 16 | 20 | 30 | 25 | 29 | 44 |
| France | 16 | 16 | 16 | 199 | 28 | 18 | 9 | 17 | 9 | 15 | 20 |
| Norway | 17 | 34 | 25 | 111 | 5 | 17 | 35 | 10 | 43 | 40 | 50 |
| Ireland | 18 | 12 | 14 | 207 | 30 | 13 | 22 | 4 | 11 | 8 | 9 |
| Austria | 21 | 38 | 33 | 158 | 39 | 23 | 10 | 51 | 28 | 18 | 21 |
| USSR (former) | 24 | 5 | 10 | 213 | 19 | 25 | 15 | 8 | 3 | 3 | 3 |
| Russian Federation | 25 | 26 | 30 | 196 | 194 | 27 | 14 | 44 | 18 | 10 | 10 |
| Philippines | 44 | 9 | 17 | 212 | 21 | 44 | 44 | 11 | 21 | 22 | 18 |
| Mexico | 66 | 3 | 5 | 215 | 7 | 67 | 52 | 1 | 1 | 1 | 1 |

Top ten countries by PageRank centrality are almost the same as top ten countries with highest migrant inflow. Germany, Poland, the USA, Turkey and Yugoslavia (former) have the highest eigenvector centrality, as they are involved in international migration by having huge migration flows and are interconnected with each other. Additionally, France is observed in top ten ranking. France has both immigrants from Germany and emigration flows to this country.

Classical centrality indices outlined countries involved in mass migration flows and countries with direct flows to or from the international migration centers, as in the previous decade. Different observations are made after the analysis of SRIC and LRIC results.

Ireland is included in the ranking by SRIC index. It is a new country in the ranking, as it was not highlighted by classical centrality indices. Ireland is a country with the population of around 3.5 million and there was an inflow from the UK of 14 200 migrants, which exceeded 0.1% of population of Ireland (share of immigrants reached 0.4%). Mexico, Turkey, the USSR (former), Poland and Yugoslavia (former) have the highest LRIC ranking, because of their interconnections with countries of huge migration flows.

An evaluation of SRIC and LRIC indices for 1989 contributed by presenting the countries with high share of immigrants (Ireland) and countries with huge emigration to popular migrants' destinations. Results of Short-Range and Long-Range Interactions Centrality indices are highly related to weighted out-degree centrality and PageRank as in the previous decade. Correlation coefficient (γ-coefficient) confirms these comparisons (Table 8).

*Table 8.* Goodman, Kruskal γ-coefficient for 1989

| | **SRIC** | **LRIC (SUM)** | **LRIC (MAX)** | **LRIC (MAXMIN)** |
|---|---|---|---|---|
| **WInDeg** | 0.697 | 0.731 | 0.719 | 0.707 |
| **WOutDeg** | 0.818 | 0.795 | 0.789 | 0.781 |
| **WDeg** | 0.82 | 0.799 | 0.792 | 0.783 |
| **WDegDiff** | –0.581 | –0.541 | –0.545 | –0.543 |
| **Clos** | 0.711 | 0.673 | 0.661 | 0.653 |
| **PageRank** | 0.699 | 0.737 | 0.723 | 0.713 |
| **Eigenvec** | 0.674 | 0.693 | 0.585 | 0.71 |
| **SRIC** | 1 | 0.892 | 0.891 | 0.883 |
| **LRIC (SUM)** | | 1 | 0.972 | 0.95 |
| **LRIC (MAX)** | | | 1 | 0.961 |
| **LRIC (MAXMIN)** | | | | 1 |

**1990–1999**

International migration in the last decade of the 20th century was characterized by the huge migrant inflows to the USA (from Mexico to Southeast Asia) and appearance of the new international migration flows from the former Soviet Union (fSU) countries and within them due to the collapse of the Soviet Union. Already established migration flows from Yugoslavia to Germany and from Turkey to Germany were still among the mass migration corridors. Additionally, Germany was the second destination country for the migrants from the fSU countries (after these countries themselves). The next group of migration flows were intra-European migration flows, predominantly from Eastern to Western European countries.

The international migration flows were the most multidimensional and massive in 1992, year after the collapse of the USSR and stabilized till the end of the decade.

In order to explore the year of the most intensive migration in the decade by analysis of the centrality indices the international migration flows in 1992 were chosen. The amount of migration flows over 50 000 of migrants in 1992 (Table 9) was the highest for the whole period from 1970 to 2013.

*Table 9.* Migration flows over 50 000 in 1992

| Origin | Destination | Migration flow |
|---|---|---|
| Russian Federation | Ukraine | 309 336 |
| Yugoslavia (former) | Germany | 267 000 |
| Mexico | USA | 213 802 |
| Ukraine | Russian Federation | 199 355 |
| Kazakhstan | Russian Federation | 183 891 |
| Poland | Germany | 143 709 |
| Uzbekistan | Russian Federation | 112 442 |
| Germany | Poland | 112 062 |
| Germany | Yugoslavia (former) | 95 720 |
| Russian Federation | Kazakhstan | 87 272 |
| Kazakhstan | Germany | 86 864 |
| Russian Federation | Germany | 84 509 |
| Turkey | Germany | 81 404 |
| Vietnam | USA | 77 735 |
| Bosnia and Herzegovina | Germany | 75 678 |
| Tajikistan | Russian Federation | 72 556 |

| Origin | Destination | Migration flow |
|---|---|---|
| Azerbaijan | Russian Federation | 69 943 |
| Romania | Germany | 67 552 |
| Kyrgyzstan | Russian Federation | 65 385 |
| Philippines | USA | 61 022 |
| Russian Federation | Belarus | 57 520 |
| Georgia | Russian Federation | 54 247 |
| Germany | Romania | 52 367 |

The ranking by centrality indices is presented in Table 10.

The ranking by weighted in-degree centrality represents not only Germany and the USA as traditional immigration countries of the previous decades, but also the Russian Federation and Ukraine. The highest emigration rate (or weighted out-degree centrality) have the Russian Federation, Germany, Yugoslavia and Kazakhstan. Germany, the USA, the Russian Federation and Ukraine had consequently the largest gross migration. The weighted degree difference centrality ranked the USA, Germany, the Russian Federation and Ukraine as countries that had largest net migration.

Countries with the highest closeness centrality index were Canada, the USA, Germany and New Zealand. Canada and New Zealand were represented because of their migration flows to and from the USA and Australia, correspondingly.

Both PageRank and eigenvector centralities rank countries based not only on their migration flows, but also account for their direct neighbors. The difference between PageRank and eigenvector centrality is that eigenvector additionally accounts for the number of migrants in migration flows. PageRank and Eigenvector results for 1992 differ considerably. PageRank apart from the USA and Germany ranked the UK, Australia and Canada, which had outflows of migrants to the USA of 40 000, 11 150 and 15 205, correspondingly. Eigenvector centrality presented Germany, Ukraine, the Russian Federation, Poland and Yugoslavia among top five countries. Poland and Yugoslavia had the outflow of migrants to Germany of 143 709 and 267 000, correspondingly.

Overall, the results estimated by classical centrality indices for the 1990s provide the ranking of countries that complies the same idea as in the previous decades. The two main groups of countries always are introduced: 1) with huge migration inflows and outflows (Germany, the USA and the Russian Federation in 1992), and 2) countries directly connected through migration flows with previous group (Canada, the UK, Kazakhstan and Yugoslavia).

*Table 10.* Rankings by centrality indices for 1992

| Country | WInDeg | WOutDeg | WDeg | WDegDiff | Clos | PageRank | EigenVec | SRIC | LRIC (SUM) | LRIC (MAX) | LRIC (MAXMIN) |
|---|---|---|---|---|---|---|---|---|---|---|---|
| Germany | 1 | 2 | 1 | 2 | 2 | 1 | 1 | 3 | 6 | 13 | 13 |
| USA | 2 | 13 | 3 | 1 | 3 | 2 | 7 | 12 | 15 | 15 | 18 |
| Russian Federation | 3 | 1 | 2 | 3 | 18 | 10 | 3 | 2 | 2 | 3 | 3 |
| Ukraine | 4 | 5 | 4 | 5 | 19 | 21 | 2 | 5 | 3 | 2 | 2 |
| Canada | 5 | 39 | 10 | 4 | 1 | 5 | 25 | 57 | 37 | 50 | 42 |
| Australia | 6 | 9 | 8 | 7 | 11 | 4 | 27 | 7 | 19 | 28 | 25 |
| UK | 7 | 8 | 7 | 24 | 22 | 3 | 11 | 8 | 10 | 16 | 23 |
| Switzerland | 8 | 23 | 12 | 6 | 6 | 9 | 13 | 64 | 55 | 56 | 64 |
| Poland | 9 | 7 | 9 | 208 | 16 | 6 | 4 | 6 | 9 | 9 | 10 |
| Netherlands | 10 | 28 | 15 | 8 | 7 | 8 | 20 | 37 | 43 | 45 | 46 |
| Yugoslavia (former) | 11 | 3 | 5 | 214 | 10 | 7 | 5 | 212 | 212 | 212 | 212 |
| Kazakhstan | 12 | 4 | 6 | 213 | 21 | 35 | 6 | 4 | 1 | 1 | 1 |
| Italy | 13 | 25 | 16 | 10 | 27 | 11 | 12 | 29 | 22 | 27 | 24 |
| Belarus | 14 | 30 | 22 | 13 | 57 | 48 | 9 | 39 | 27 | 17 | 21 |
| Croatia | 15 | 48 | 26 | 9 | 26 | 23 | 18 | 26 | 24 | 23 | 19 |
| Romania | 16 | 17 | 17 | 194 | 24 | 15 | 8 | 34 | 11 | 12 | 12 |
| Turkey | 17 | 11 | 13 | 205 | 23 | 14 | 10 | 11 | 5 | 6 | 6 |
| Sweden | 18 | 50 | 32 | 12 | 8 | 18 | 40 | 18 | 59 | 73 | 73 |
| New Zealand | 20 | 31 | 25 | 21 | 9 | 12 | 51 | 17 | 29 | 29 | 26 |
| Denmark | 22 | 46 | 33 | 16 | 4 | 13 | 36 | 53 | 64 | 71 | 66 |
| Uzbekistan | 24 | 10 | 14 | 210 | 55 | 64 | 15 | 13 | 4 | 4 | 4 |
| Norway | 28 | 66 | 47 | 17 | 5 | 22 | 56 | 25 | 74 | 86 | 83 |
| Azerbaijan | 34 | 19 | 24 | 201 | 132 | 72 | 23 | 30 | 14 | 10 | 9 |
| Bosnia and Herzegovina | 47 | 12 | 19 | 211 | 28 | 43 | 41 | 14 | 7 | 7 | 8 |
| Tajikistan | 50 | 18 | 27 | 206 | 136 | 106 | 37 | 28 | 13 | 8 | 7 |
| Philippines | 62 | 16 | 23 | 209 | 29 | 49 | 64 | 9 | 20 | 22 | 22 |
| Vietnam | 64 | 14 | 20 | 212 | 25 | 58 | 47 | 10 | 17 | 19 | 17 |
| Mexico | 82 | 6 | 11 | 215 | 13 | 77 | 76 | 1 | 8 | 5 | 5 |

The estimation of the Short- and Long-Range Interactions Centrality indices provides different from the previous one list of countries.

SRIC ranking presents mostly the emigration countries. This fact is also confirmed by the Goodman-Kruskal correlation coefficient (Table 11) between SRIC and weighted out-degree indices (0.8). LRIC indices rank countries of the fSU in top four: Kazakhstan, Uzbekistan, The Russian Federation and Ukraine. Other emigration countries are also considered by these indices.

Comparison of the classical centrality indices and the SRIC and LRIC results lead to a conclusion that Short-Range and Long-Range Interactions Centrality indices were highly related to the weighted out-degree and weighted degree centralities $\gamma \approx (0.8)$ and compared to PageRank and eigenvector, they outline additionally countries with intense emigration flows, not only the countries connected to the attractive migrants' destinations.

*Table 11.* Goodman, Kruskal $\gamma$-coefficient for 1992

| | **SRIC** | **LRIC (SUM)** | **LRIC (MAX)** | **LRIC (MAXMIN)** |
|---|---|---|---|---|
| **WInDeg** | 0.669 | 0.719 | 0.698 | 0.714 |
| **WOutDeg** | 0.818 | 0.84 | 0.832 | 0.793 |
| **WDeg** | 0.812 | 0.848 | 0.823 | 0.797 |
| **WDegDiff** | –0.426 | –0.421 | –0.427 | –0.386 |
| **Clos** | 0.642 | 0.643 | 0.624 | 0.607 |
| **PageRank** | 0.638 | 0.673 | 0.647 | 0.675 |
| **Eigenvec** | 0.64 | 0.708 | 0.694 | 0.681 |
| **SRIC** | 1 | 0.873 | 0.864 | 0.841 |
| **LRIC (SUM)** | | 1 | 0.957 | 0.924 |
| **LRIC (MAX)** | | | 1 | 0.919 |
| **LRIC (MAXMIN)** | | | | 1 |

**2000–2013**

The last period of the international migration presented in our database is from 2000 to 2013. The major international migration flows occurred between the following groups of countries. First, the migration flow from Mexico, the Philippines and Vietnam to the USA were still of considerable level. Second, new Asian countries, India and China, appeared among labor force suppliers for the USA. The next destination of migrants from the developing countries was Spain. The immigrants from Ecuador, Morocco, Colombia and Argentina were moving to Spain till the beginning of the economic crisis in 2008. After 2008 Spain is no longer attractive for immigrants due to the high level of unemployment and becomes an emigration country [15]. Flows between the fSU countries were diminishing after 2007 and migration from the Russian Federation and Kazakhstan to Germany was decreasing accordingly. According to Eurostat statistics [37] Greece was one of the countries that experienced the highest growth in number of international migrants in recent time. Also it was among the countries highly involved in migration in 1972 and was ranked by LRIC indices in top 10 countries. However, since 1998 the databases [32, 33] do not contain regular statistics on migration flows for Greece, that is why Greece is not presented in rankings by centrality measures.

These processes in international migration in the last decade lead to the development of the following migration flows over 50 000 (Table 12) in 2013.

*Table 12.* Migration flows over 50 000 in 2013

| Origin | Destination | Migration flow |
|---|---|---|
| Mexico | USA | 135 028 |
| China | USA | 71 798 |
| Spain | Romania | 70 055 |
| India | USA | 68 458 |
| Romania | Italy | 59 347 |
| Philippines | USA | 54 446 |

The considerable reduction in the number of international migration flows over 50 000 is observed compared to 1992. Centrality indices reflect these changes noticeably (Table 13).

From the results for weighted in-degree centrality we can conclude that the highest number of immigrants were received by the USA, Italy and the UK. According to the ranking by weighted out-degree, Spain, India and China had the highest migrant out-flow. Weighted degree ranking highlights the USA, Spain, Italy and the UK, which had the greatest gross migration rate. The weighted degree difference or the highest net migration flow was in the USA, Canada, the UK and Italy.

Different results can be obtained from the estimation of the level of closeness: the USA is still the first, however, Mexico, Netherlands, Spain and Switzerland are presented. These countries had intense migration in-flows (the USA) or out-flows (Spain) itself, or had migration flows to or from the countries with intense migration [26]. Mexico-US migration route was established historically, and now Mexicans are accounted for 28% of foreign-born population in the USA [36]. Netherlands and Switzerland were connected through migration flows to Italy, which was the second immigration country after the USA.

Eigenvector and PageRank highlight the "rich-club" group of countries: the USA, Italy, the UK and Spain. These countries are involved in the process of migration more than others and in addition had flows between each other. In this case eigenvector and PageRank centralities can show how "mobile" is the population of countries.

*Table 13.* Rankings by centrality indices for 2013

| Country | WInDeg | WOutDeg | WDeg | WDegDiff | Clos | PageRank | EigenVec | SRIC | LRIC (SUM) | LRIC (MAX) | LRIC (MAXMIN) |
|---|---|---|---|---|---|---|---|---|---|---|---|
| USA | 1 | 19 | 1 | 1 | 1 | 1 | 2 | 22 | 6 | 10 | 10 |
| Italy | 2 | 5 | 3 | 4 | 6 | 6 | 4 | 11 | 10 | 11 | 16 |
| UK | 3 | 10 | 4 | 3 | 30 | 3 | 1 | 9 | 9 | 4 | 7 |
| Canada | 4 | 44 | 5 | 2 | 10 | 7 | 12 | 74 | 37 | 43 | 30 |
| Spain | 5 | 1 | 2 | 215 | 3 | 2 | 3 | 1 | 1 | 1 | 1 |
| Switzerland | 6 | 12 | 7 | 6 | 5 | 5 | 6 | 35 | 44 | 54 | 80 |
| Netherlands | 7 | 8 | 8 | 10 | 4 | 8 | 11 | 17 | 14 | 23 | 27 |
| Sweden | 8 | 21 | 15 | 5 | 9 | 11 | 19 | 15 | 30 | 38 | 35 |
| Belgium | 9 | 14 | 10 | 9 | 7 | 12 | 9 | 23 | 19 | 28 | 45 |
| Romania | 10 | 6 | 6 | 198 | 14 | 17 | 5 | 2 | 2 | 2 | 2 |
| Germany | 11 | 11 | 9 | 23 | 37 | 10 | 7 | 12 | 4 | 8 | 9 |
| New Zealand | 12 | 16 | 13 | 14 | 8 | 4 | 14 | 5 | 23 | 15 | 15 |
| France | 13 | 9 | 12 | 192 | 36 | 15 | 8 | 7 | 3 | 5 | 5 |
| Norway | 14 | 52 | 23 | 7 | 11 | 16 | 23 | 32 | 45 | 49 | 24 |

| Country | WInDeg | WOutDeg | WDeg | WDegDiff | Clos | PageRank | EigenVec | SRIC | LRIC (SUM) | LRIC (MAX) | LRIC (MAXMIN) |
|---|---|---|---|---|---|---|---|---|---|---|---|
| Australia | 15 | 31 | 22 | 8 | 33 | 9 | 20 | 18 | 21 | 25 | 21 |
| Morocco | 18 | 17 | 20 | 166 | 31 | 21 | 10 | 8 | 7 | 7 | 6 |
| Poland | 23 | 13 | 18 | 210 | 44 | 20 | 28 | 4 | 8 | 6 | 4 |
| India | 32 | 2 | 11 | 214 | 24 | 26 | 32 | 3 | 5 | 3 | 3 |
| Mexico | 45 | 4 | 14 | 212 | 2 | 56 | 40 | 49 | 40 | 35 | 65 |
| Philippines | 53 | 7 | 19 | 211 | 28 | 48 | 51 | 16 | 16 | 12 | 11 |
| China | 73 | 3 | 16 | 213 | 23 | 55 | 90 | 6 | 11 | 9 | 8 |
| Syrian Arab Republic | 133 | 37 | 49 | 200 | 61 | 137 | 137 | 10 | 28 | 27 | 22 |

Ranking by classic centrality indices provided us with the information about countries with the highest in- and out-flows of migrants, net migration flow, level of closeness to huge migration flows and countries most involved in migration process. Short-Range and Long-Range Interactions Centralities can help us to explore the international migration network from the different perspective.

Spain, Romania, India and Poland had the highest ranks according to the index of Short-Range Interactions Centralities. These results are highly related to the weighted out-degree. Additionally, SRIC accounts for the first-order indirect interactions and the population of destination country. That is why there was a little change in the order of countries with intense emigration flows.

Three of LRIC indices show almost similar results: Spain, Romania, France, Germany, Poland and India are at the top of rankings. Spain has the highest emigration rate. Romania, India and France have the migration flows to countries with huge population and intense migration flows. There was a huge flow from India to the USA, the USA has large population and is a popular country of migrants' destination [35]. France is presented in ranking by LRIC indices, because it has migration flows to Spain (10 548) and to the UK (24 313). Romania also had migration flows to the UK. Poland did not appear among countries with highest emigration rate (weighted out-degree), how-ever, it had migration flow of almost 10000 migrants to Norway with population of around 5 million people. The share of this migrant inflow (0.2%) exceeded 0.1% of the population of Norway. This result is important to be considered as when migration flow is more than level expected by the destination country, it can lead to negative consequences for both migrants and the population of destination country.

The results introduced by classical centralities and SRIC, LRIC indices both outline the emigration countries. However, SRIC and LRIC indices introduce additionally the emigration countries with considerable for the population of destination country share of migrants (Poland).

*Table 14.* Goodman, Kruskal γ-coefficient for 2013

| | SRIC | LRIC (SUM) | LRIC (MAX) | LRIC (MAXMIN) |
|---|---|---|---|---|
| **WInDeg** | 0.716 | 0.746 | 0.716 | 0.742 |
| **WOutDeg** | 0.839 | 0.793 | 0.774 | 0.742 |
| **WDeg** | 0.831 | 0.798 | 0.798 | 0.742 |
| **WDegDiff** | –0.414 | –0.359 | –0.359 | –0.341 |
| **Clos** | 0.704 | 0.69 | 0.69 | 0.642 |
| **PageRank** | 0.716 | 0.76 | 0.76 | 0.714 |
| **Eigenvec** | 0.705 | 0.729 | 0.704 | 0.676 |
| **SRIC** | 1 | 0.845 | 0.845 | 0.799 |
| **LRIC (SUM)** | | 1 | 0.934 | 0.885 |
| **LRIC (MAX)** | | | 1 | 0.9 |
| **LRIC (MAXMIN)** | | | | 1 |

The main goal of the analysis of each decade provided above was to represent major migration flows in terms of the network analysis by introducing the ranking of countries based on the centrality indices. Overall, the analysis of classical and SRIC, LRIC centralities by decades introduced the following outcomes. First, classical centrality indices analysis outline the occurrence and development of the major international migration flows in each decade. Second, SRIC and LRIC outline the influence of these flows on the population on destination countries and changes in interconnections in international migration network.

### *5.2. Centrality indices in dynamics*

After the analysis of the centrality indices for each decade was made, it is essential to aggregate the results by each index for the whole period and observe the dynamics of the centrality indices. For this purpose the diagrams with the dynamics of each centrality index for around ten countries are constructed and the description of the major peaks is given. Centrality indices are presented in the same order as in the previous Section. First, classical centrality indices are provided and then the dynamics of Short and Long-Range Centrality indices is illustrated.

The dynamics of weighted in-degree centrality or the migrant inflow (Figure 2 and 3) shows countries with interesting changes in this index over the period. The lines that correspond to the USA, Germany and the Russian Federation had remarkable changes over the period. First, the peak for the US of more than 1.8 million of migrants is explained by the increase of migrant inflow from Mexico from 1986 to 1989. Interestingly, this immigration is related to the immigration amnesty made by the USA in the 1986, which legalized immigrants already resided in the US and is not related to the arrival of new people. Second, immigration to Germany reached its maximum (1.6 million immigrants) in 1990, which is explained by the reunification of Germany. The second peak of 1.4 million immigrants in 1992 relates to the dissolution of the Soviet Union and the inflows from the Russian Federation, Ukraine, Kazakhstan and other countries of the fSU. The third country with outstanding dynamics of this index is the Russian Federation. The explanation of the peak after 1991 is the same: huge immigration rate from the fSU countries. Additionally, Germany and the Russian Federation were the only countries in this list with declining in immigration flows from 1970 to 2013.

Graphs of three countries (the USA, Germany and the Russian Federation) mentioned above were outlined from Figure 2 to have better representation of the dynamics for other countries (Figure 3). On this Figure European countries and Canada are presented. The common characteristic of the dynamics of migrant inflow to almost all European countries is the downward tendency in 2008 because of the economic crisis. This fact had the highest influence on immigration rate to Spain and lead to the drop of almost 400 thousands migrants. Overall migration inflow to almost all countries presented in Figure 2 and 3 (except Germany and the Russian Federation) developed over the period and varied more considerably from the beginning of the 20$^{th}$ century.

The dynamics of weighted out-degree centrality or the migrant outflow is represented also by two graphs (Figure 4 and 5). Countries with the most no-ticeable changes in this centrality index over the period are shown in Figure 4. The major peak of emigration for five countries occurred between 1989 and 1994. Migration from Mexico to the USA in 1989 due to immigration amnesty mentioned above shows the hugest peak in weighted out-degree centrality. The next cause of the largest peaks in this period – fall of the Soviet Union influenced emigration of around 700 thousand migrants from the Russian Federation in 1992 and from 200 to 300 thousand migrants from Ukraine from 1991 to 1994. Another emigration wave in 1992 was from former Yugoslavia because of military conflicts occurred in this period.

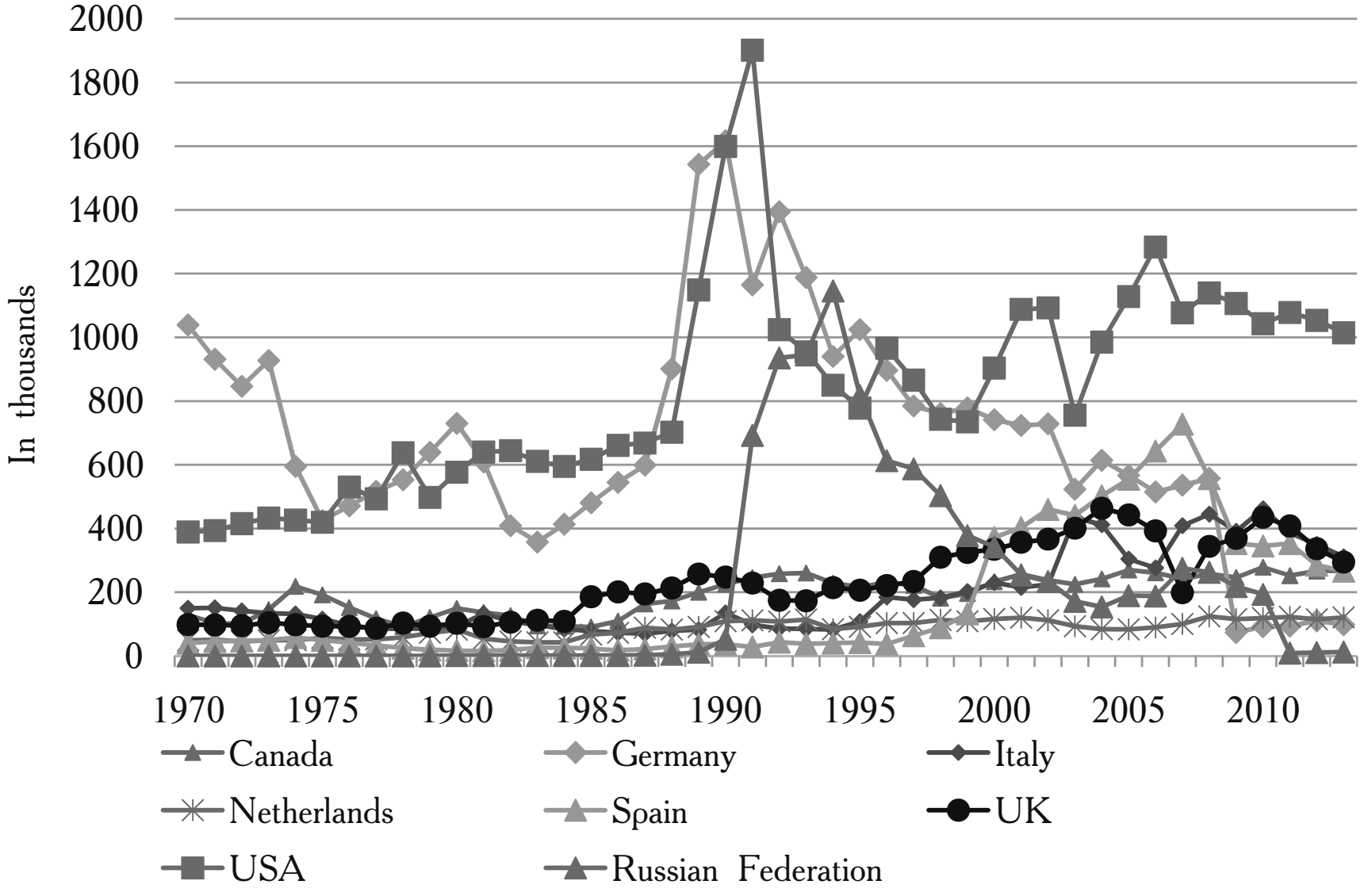

**Fig. 2.** Weighted In-degree (Migrant inflow) 1

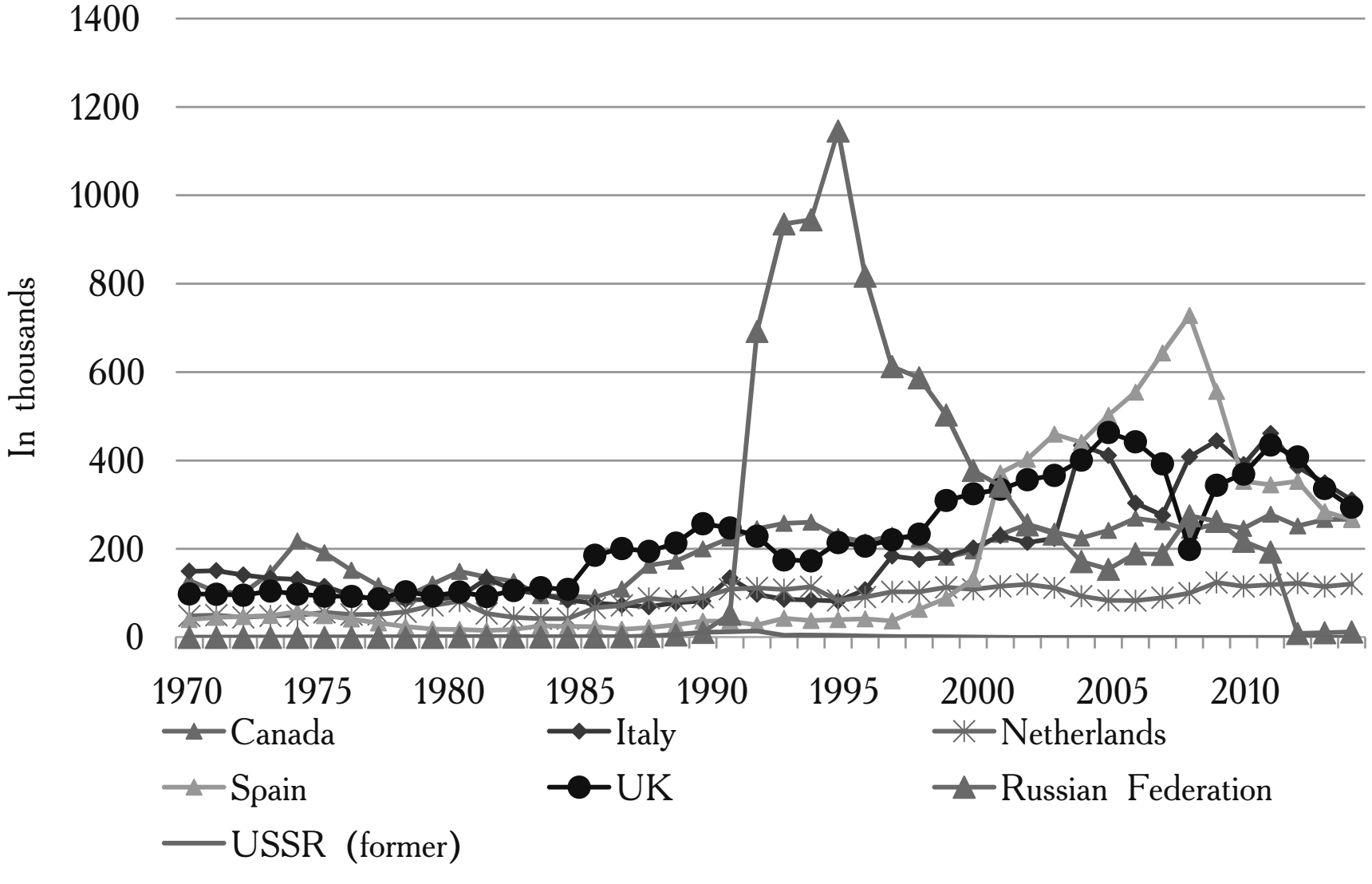

**Fig. 3.** Weighted in-degree (Migrant inflow) 2

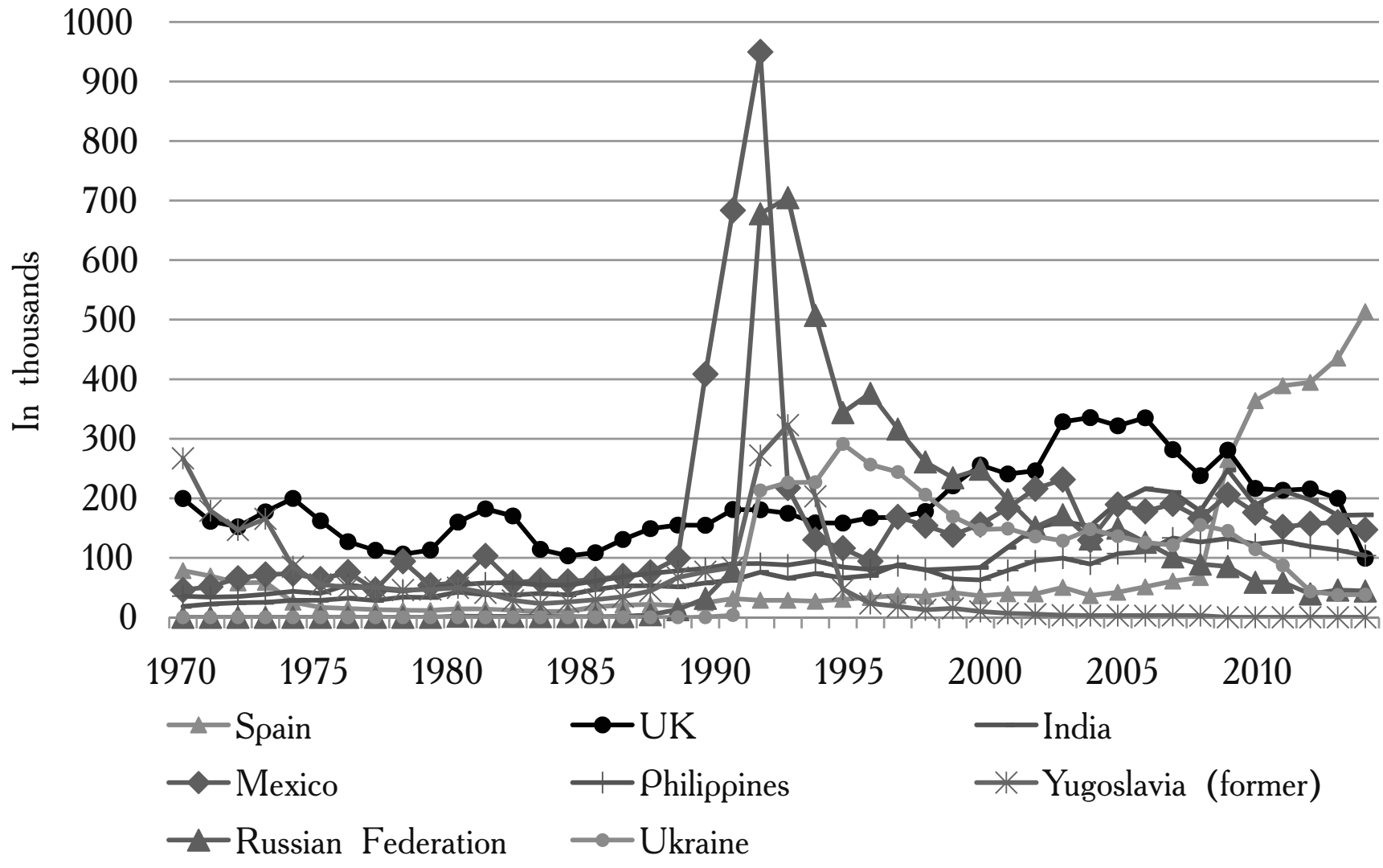

**Fig. 4.** Weighted Out-degree (Migrant outflow) 1

In order to explore the changes in less massive peaks of emigration happened in the 2000s Figure 5 with the dynamics of weighted out-degree centrality is presented. The rise of weighted out-degree of Spain in 2008 is explained by the change from the high immigration to huge emigration rate of Spain in after the economic crisis. On the contrary outflows from other countries presented on this graph declined after 2008. Overall, for the outflows of last decade economic crisis in 2008 was the main reason of changes in the rate of emigration.

As weighted degree and weighted degree difference are gross and net migration flows correspondingly, their dynamics follow the same tendencies as the dynamics of weighted in-degree and weighted out-degree centralities. Graphs of dynamics of these indices are shown in the Appendix.

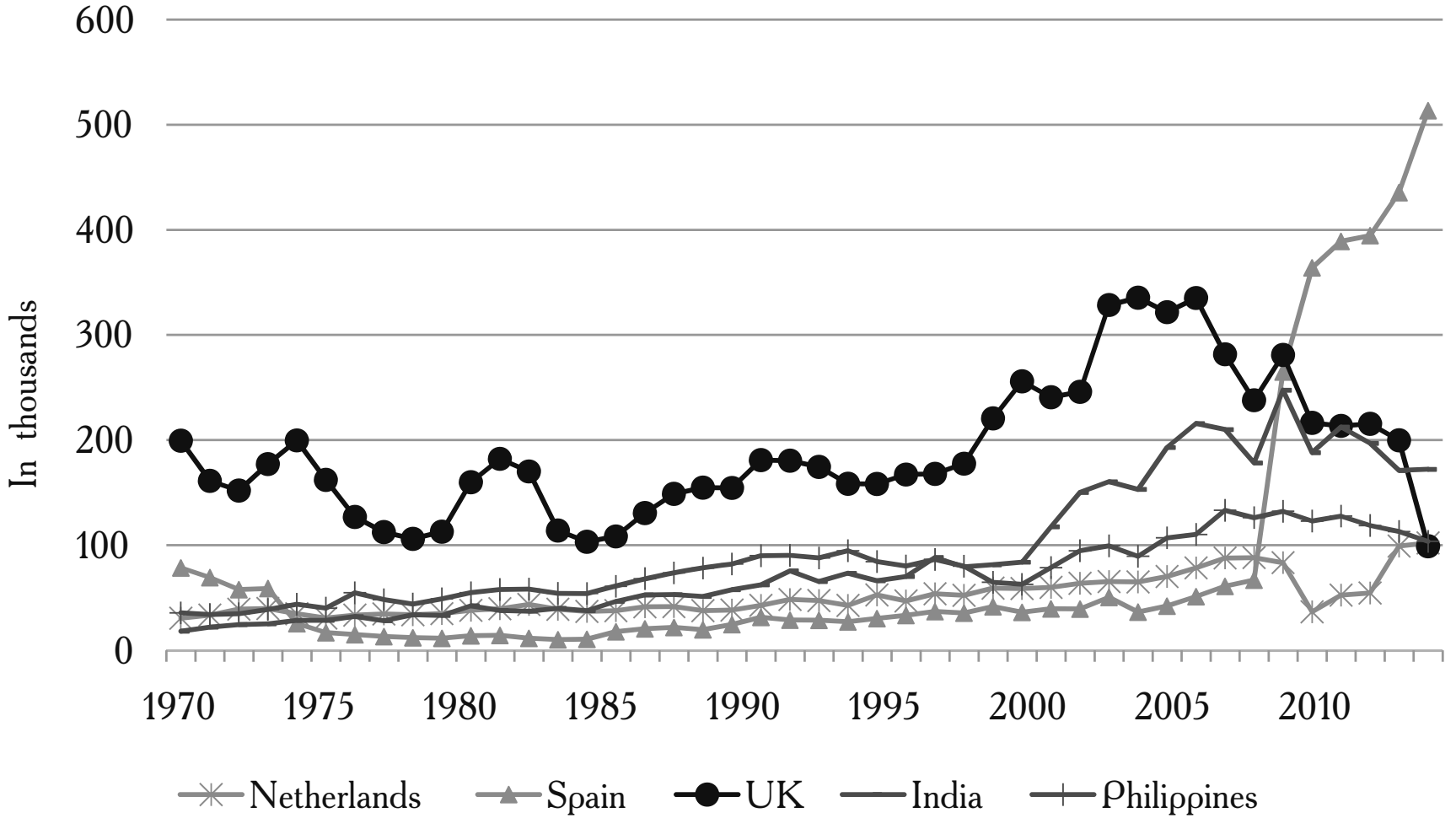

**Fig. 5.** Weighted Out-degree (Migrant outflow) 2

After the overview of dynamics of migration inflows and outflows was made it is interesting to observe the changes in the degree of importance of countries based on PageRank centrality. The countries with the highest rank by this centrality over the period are shown on Figure 6. The dynamics of this index for the majority of countries was quite stable over the period. However, the PageRank centrality of Germany had a noticeable decline from 1970 to 2013. The reason is probably that Germany was constantly involved into the process of mass migration in the first two decades and from the middle 1990s flows to and from Germany began to fall down.

Short-Range and Long-Range Interactions Centralities' dynamics have noticeable differences from classical indices. Note that SRIC and LRIC values may vary from 0 to 1.

Figure 7 presents the SRIC dynamics of the countries that were leaders in ranking by this index over the period. We observe the changes occurred in the international migration network based on the Short-Range Interactions. From 1970 till 1989 Germany, Turkey and Poland had the highest SRIC values. The same fact was observed in the analysis of indices by decades. The peak of emigration from Mexico to the USA in 1989 is presented on this graph as well. The migration flows after the break-up of the Soviet Union are shown in SRIC dynamics. Russian Federation, Kazakhstan, Germany – countries involved in the migration flows from 1990 till the beginning of 2000s have the highest

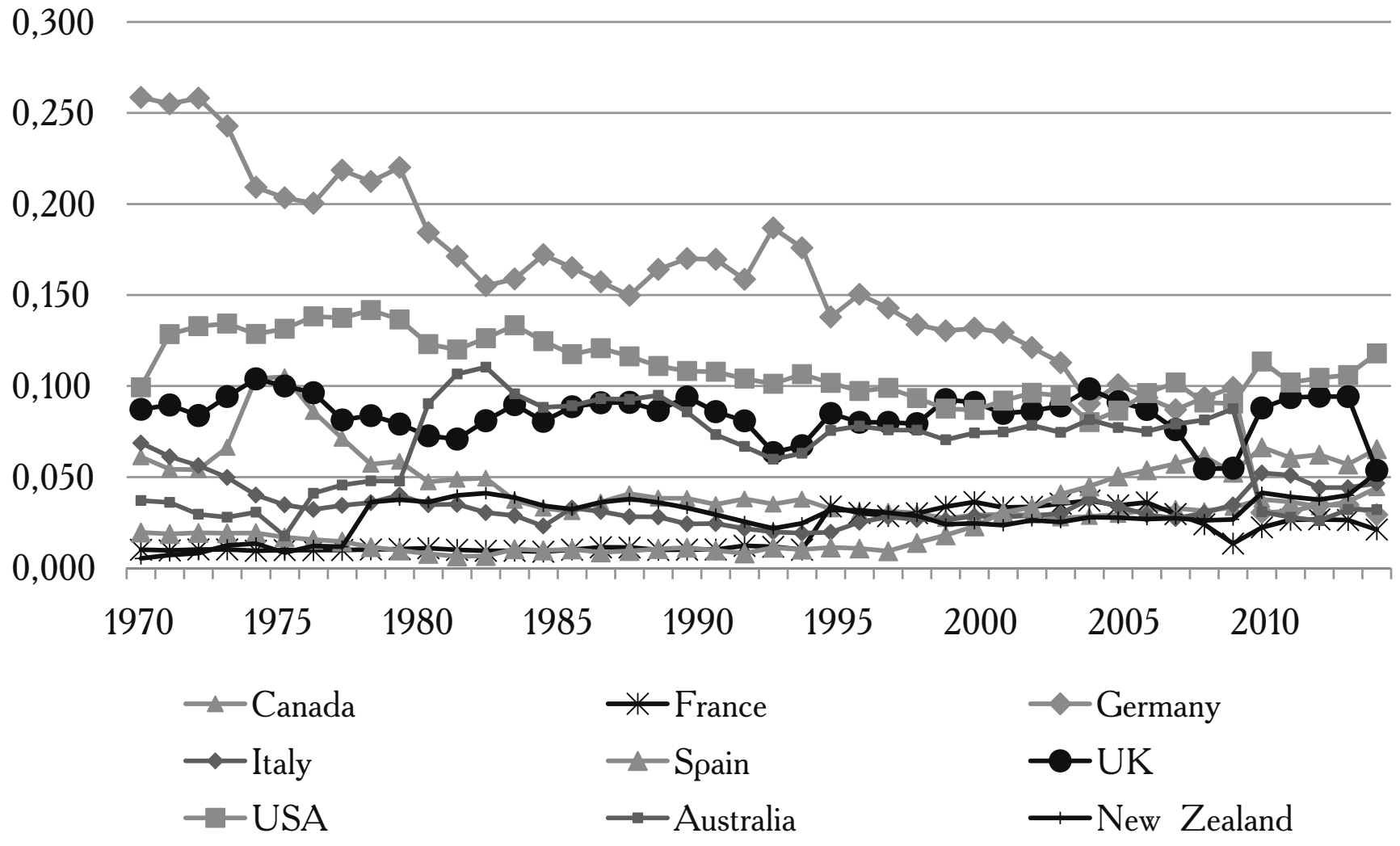

**Fig. 6.** PageRank

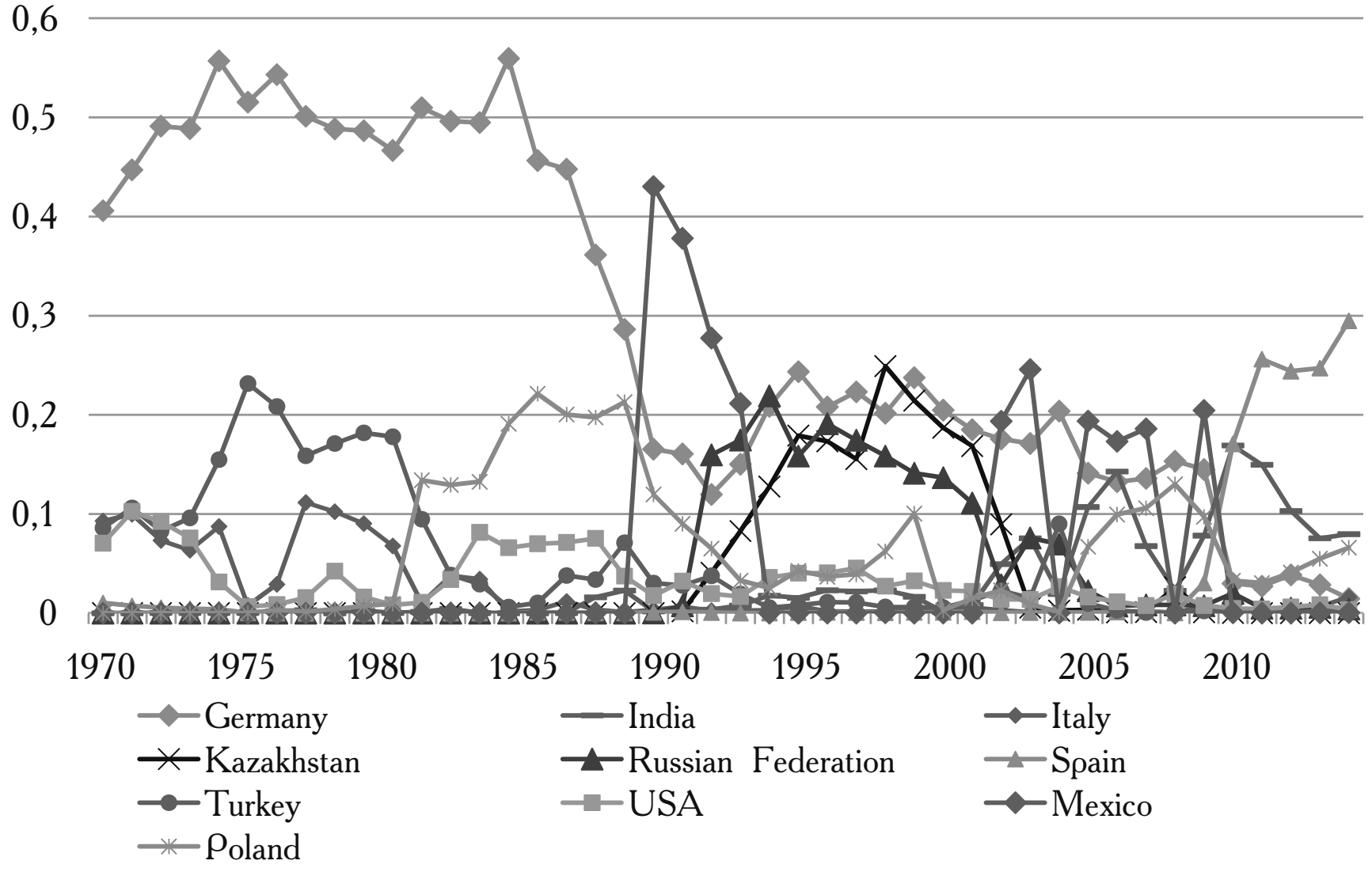

**Fig. 7.** SRIC

SRIC index for the corresponding period. The last decade is characterized by the strengthening of influence of Asian countries (India, the Philippines), because of the emigration to the USA from them and the upward trend for Spain because of both emigration and immigration from and to this country. Overall, the dynamics of SRIC index illustrates the development in influence of countries most involved in process of international migration and most interconnected with them.

In Figure 8 LRIC (SUM) dynamics is presented. Other versions of LRIC indices are shown in the Appendix. The major picture of the dynamics of LRIC indices is the same as for SRIC index. However, its values are low and vary from 0 to 0.16 even for the countries with highest ranking by this centrality index.

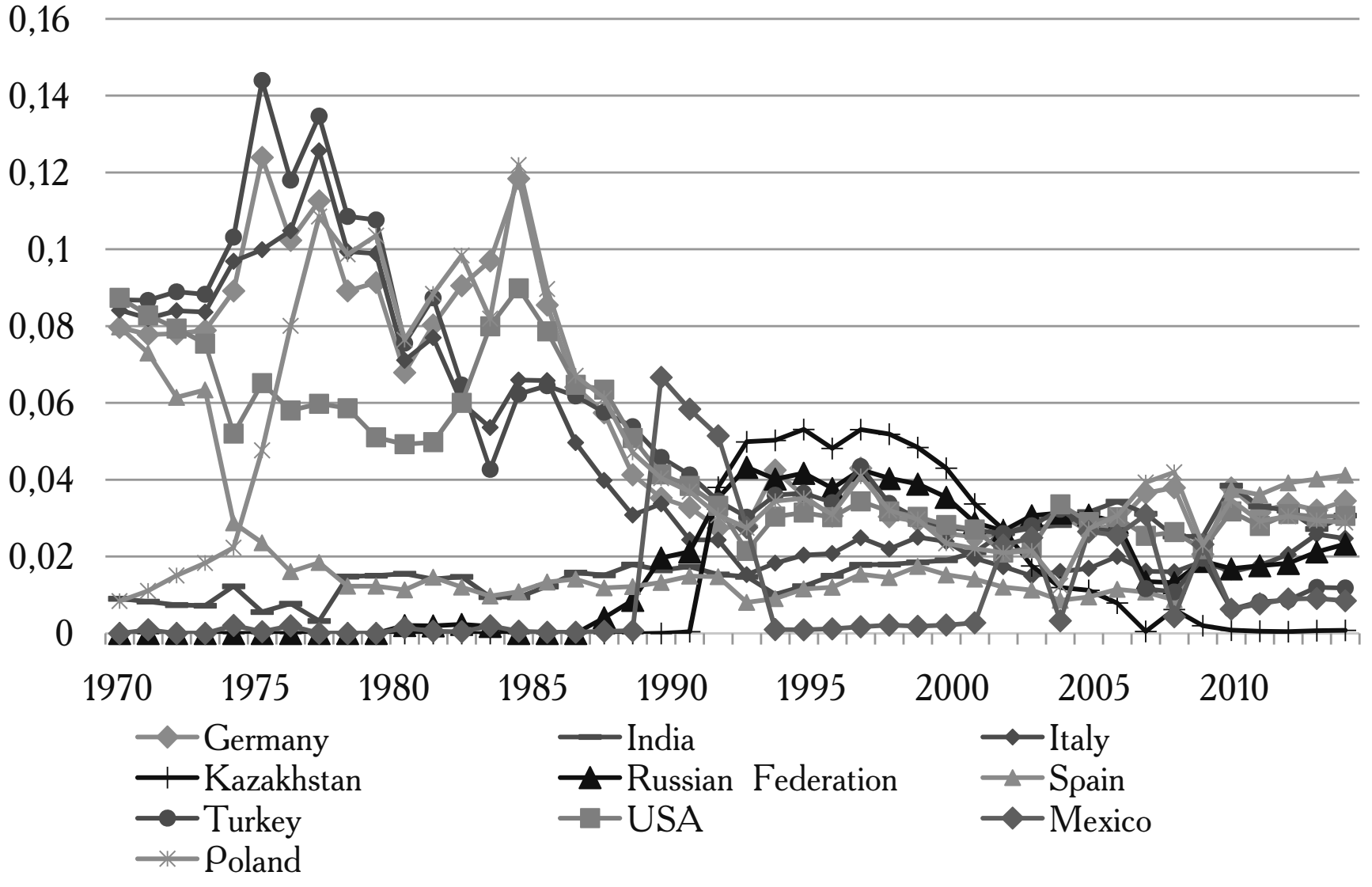

**Fig. 8.** LRIC (SUM)

Another difference from SRIC dynamics is that the peaks for Turkey and Kazakhstan were more considerable. The reason of this rise can be explained by the increase in their influence by reaching certain share of population of destination countries and in strengthening of interconnections with attractive countries for migrants. LRIC of Poland was escalating till 1976 simultaneously with the rise of the level of emigration from Poland to Germany. They follow the same trend of LRIC (MAX) index. In the last decade there was no-ticeable upward trend of LRIC for Spain, the same fact was observed by SRIC and weighted out-degree centrality.

To sum up, LRIC dynamics had changes in the dynamics for certain countries related to the rise of emigration from them to the countries-centers of immigration or the increase in overall emigration level.

The dynamics of classical centrality indices and Short and Long-Range Interactions Centralities shows the dynamics of international migration from the different prospective. The weighted degree centralities shows the dynamics of international migration itself. PageRank centrality illustrates the changes related to the number of interconnections with other countries. LRIC and SRIC indices present the dynamics related to the changes in the level of emigration from the given country and in the level of interconnections with the attractive for immigrants countries.

## Conclusion

International migration can be modeled in various ways. Extensive amount of works study the international migration flows on country-to-country level and analyze the causes of their emerging. Network analysis allows to represent all countries as a system and consider the migration flows between any two countries as an imprescriptible part of the international migration flows in the whole network.

Estimation of classical centrality indices is the one of the possible ways to analyze countries' influence in the network through migration flows. Our work goes a step further and allows to consider indirect connections of countries in the international migration network and a node attribute – the population of destination country. This idea is implied through the Short-Range and Long-Range Interactions Centralities.

The analysis are applied to annual data on migration flows, the results of the estimations have been compared for each decade and the dynamics of indices are presented. Our methodology outline not only the countries with large number of immigrants or emigrants, but also the countries with migrant outflows considerable for the population of destination country and emigration to the popular destination countries. These results are important in order to provide countries highly involved in the process of international migration with relevant migration policy.

# 7. Appendix

## *7.1. Database description*

*Table 15.* The list of countries provided the statistics for the 2008 Revision

| Country | Immigration | Emigration | Country | Immigration | Emigration |
|---|---|---|---|---|---|
| Australia | 1976–2008 | 1976–2008 | Italy | 1986–2006 | 1986–2006 |
| Austria | 1996–2008 | 1996–2008 | Latvia | 1995–2008 | 1995–2008 |
| Belgium | 1997–2007 | 1970–2007 | Lithuania | 2001–2008 | 2001–2008 |
| Canada | 1970–2008 | – | Luxembourg | 1980–2007 | 1980–2007 |
| Croatia | 1992–2008 | 1992–2008 | Netherlands | 1970–2007 | 1970–2007 |
| Czech Republic | 1993–2007 | 1993–2007 | New Zealand | 1979–2008 | 1979–2008 |
| Denmark | 1980–2008 | 1980–2008 | Norway | 1980–2008 | 1980–2008 |
| Estonia | 2004–2007 | 2004–2007 | Poland | 1999–2008 | 1999–2008 |
| Finland | 1980–2008 | 1980–2008 | Slovakia | 1993–2008 | 1993–2008 |
| France | 1994–2007 | – | Slovenia | 1996–2007 | 1996–2007 |
| Germany | 1970–2007 | 1970–2007 | Spain | 1983–2008 | 2002–2008 |
| Hungary | 1995–2007 | 1995–2007 | Sweden | 1970–2008 | 1970–2008 |
| Iceland | 1986–2008 | 1986–2008 | Switzerland | 1991–2007 | 1991–2007 |
| Israel | 1995–2008 | – | UK | 1970–2007 | 1970–2007 |
| | | | USA | 1970–2008 | – |

*Table 16.* The list of countries provided the statistics for the 2015 Revision

| Country | Immigration | Emigration | Country | Immigration | Emigration |
|---|---|---|---|---|---|
| Armenia | 2000–2009 | 2000–2009 | Latvia | 1990–2009 | 1995–2009 |
| Australia | 1980–2008 | 1980–2008 | Liechtenstein | 1999, 2008–2013 | 1999, 2010–2013 |
| Austria | 1996–2008, 2010–2012 | 1996–2008, 2010–2012 | Lithuania | 1992–2013 | 1998–2013 |
| Azerbaijan | 1995–2009 | 1995–2009 | Luxembourg | 1998–2013 | 1980–2013 |
| Belarus | 2000–2009 | 2000–2009 | Malta | 1997–2001, 2007 | 1997–2001, 2007 |

| Country | Immigration | Emigration | Country | Immigration | Emigration |
|---|---|---|---|---|---|
| Belgium | 1985–2013 | 2010–2013 | Netherlands | 1980–2013 | 1980–2013 |
| Canada | 1980–2013 | – | New Zealand | 1980–2013 | 1980–2013 |
| Croatia | 1992–2013 | 1991–2013 | Norway | 1980–2013 | 1980–2013 |
| Cyprus | 1997–2007 | 2002–2008 | Poland | 1999–2009 | 1999–2008 |
| Czech Republic | 1994–2007 | 1993–2007 | Portugal | 1992–2008 | – |
| Denmark | 1980–2013 | 1980–2013 | Republic of Moldova | 1993–2010 | – |
| Estonia | 1992–2013 | 1998–2013 | Romania | 1994–2013 | 1990–2013 |
| Finland | 1980–2013 | 1980–2013 | Russian Federation | 1991–2010 | 1991–2010 |
| France | – | – | Slovakia | 1991–2013 | 1991–2013 |
| Germany | 1980–2008 | 1980–2008 | Slovenia | 1990–2013 | 1996–2013 |
| Country | Immigration | Emigration | Country | Immigration | Emigration |
| Greece | 1985–1993, 1995–1998, 2007–2008 | – | Spain | 1985–2013 | 2002–2013 |
| Hungary | 1995–2013 | 1995–2013 | Sweden | 1985–2013 | 1980–2013 |
| Iceland | 1986–2013 | 1986–2013 | Switzerland | 1991–2013 | 1991–2013 |
| Ireland | 2006–2013 | 1987–2013 | The former Yugoslav Republic of Macedonia (TfYR of Macedonia) | 1996, 2000–2002, 2004–2008, 2011–2012 | 2000, 2004–2008, 2011–2012 |
| Italy | 1985–2013 | 1986–2013 | Ukraine | 2000–2006 | 2000–2006 |
| Kazakhstan | 2000–2009 | 2000–2009 | UK | 1985–2013 | 1998–2006, 2002–2012 |
| Kyrgyzstan | 1990–2008 | 1990–2008 | USA | 1980–2013 | – |

*Table 17.* Examples of inconsistent observations

| Origin | Destination | Year | Min | Max | Max/Min ratio |
|---|---|---|---|---|---|
| Norway | UK | 1994 | 1 | 1000 | 1000 |
| Russian Federation | UK | 1995 | 1 | 1000 | 1000 |
| Czech Republic | Armenia | 2003 | 1 | 1137 | 1137 |
| Latvia | Canada | 2012 | 1 | 1588 | 1588 |
| Slovakia | Canada | 2011 | 1 | 1776 | 1776 |
| Slovakia | TfYR of Macedonia | 2000 | 1 | 2654 | 2654 |
| Estonia | USA | 2012 | 1 | 3748 | 3748 |
| Norway | UK | 1993 | 1 | 4000 | 4000 |
| Belgium | USA | 2010 | 1 | 4003 | 4003 |
| Ireland | Ireland | 2006 | 2 | 2393 | 1196.5 |
| Croatia | Sweden | 2007 | 2 | 2457 | 1228.5 |
| Italy | Germany | 2000 | 2 | 3441 | 1720.5 |
| Kazakhstan | TfYR of Macedonia | 1996 | 2 | 3805 | 1902.5 |
| Kazakhstan | Hungary | 2011 | 3 | 5804 | 1934.67 |
| Spain | Hungary | 2010 | 3 | 6581 | 2193.67 |
| Belarus | Russian Federation | 1994 | 3 | 15359 | 5119.67 |
| Denmark | Romania | 2007 | 4 | 4019 | 1004.75 |
| Belarus | Russian Federation | 1992 | 4 | 6650 | 1662.50 |
| Latvia | Russian Federation | 1993 | 6 | 11375 | 1895.83 |
| Iceland | Russian Federation | 1996 | 6 | 15137 | 2522.83 |
| Croatia | Russian Federation | 1995 | 6 | 17202 | 2867 |

### *7.2. Centrality indices in dynamics*

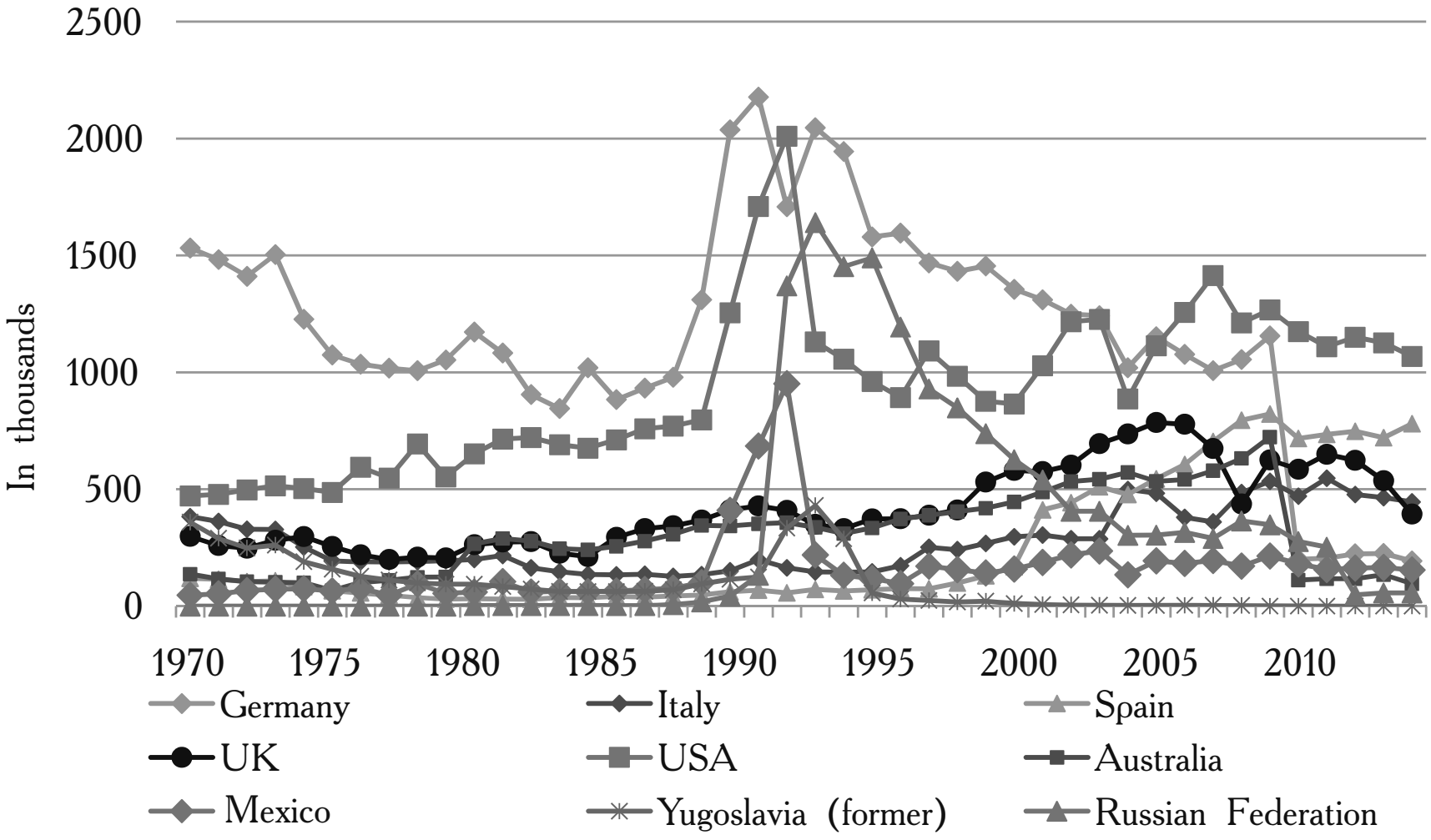

**Fig. 9.** Weighted Degree (Gross migration flow)

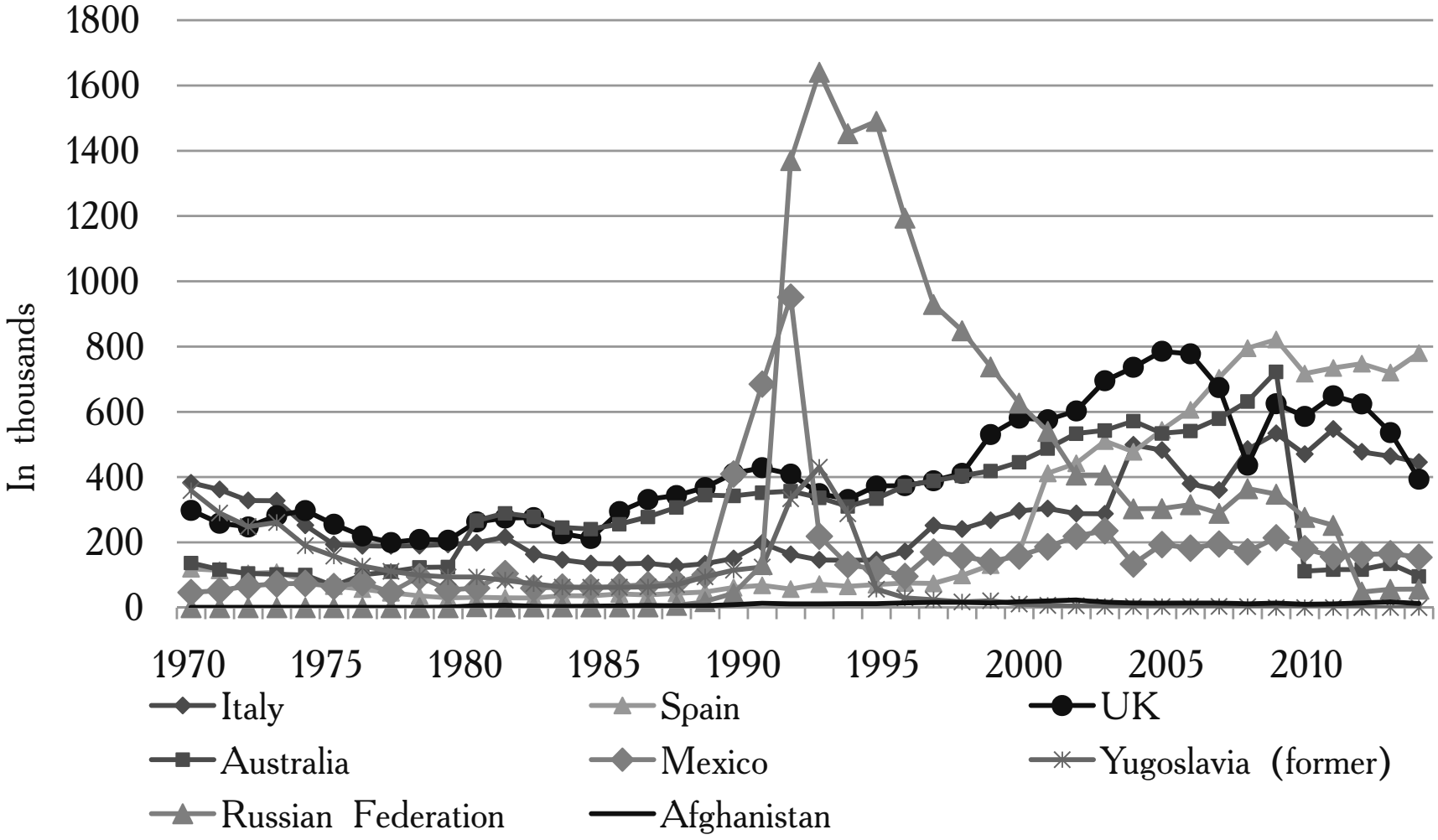

**Fig. 10.** Weighted Degree (Gross migration flow) without Germany and the USA

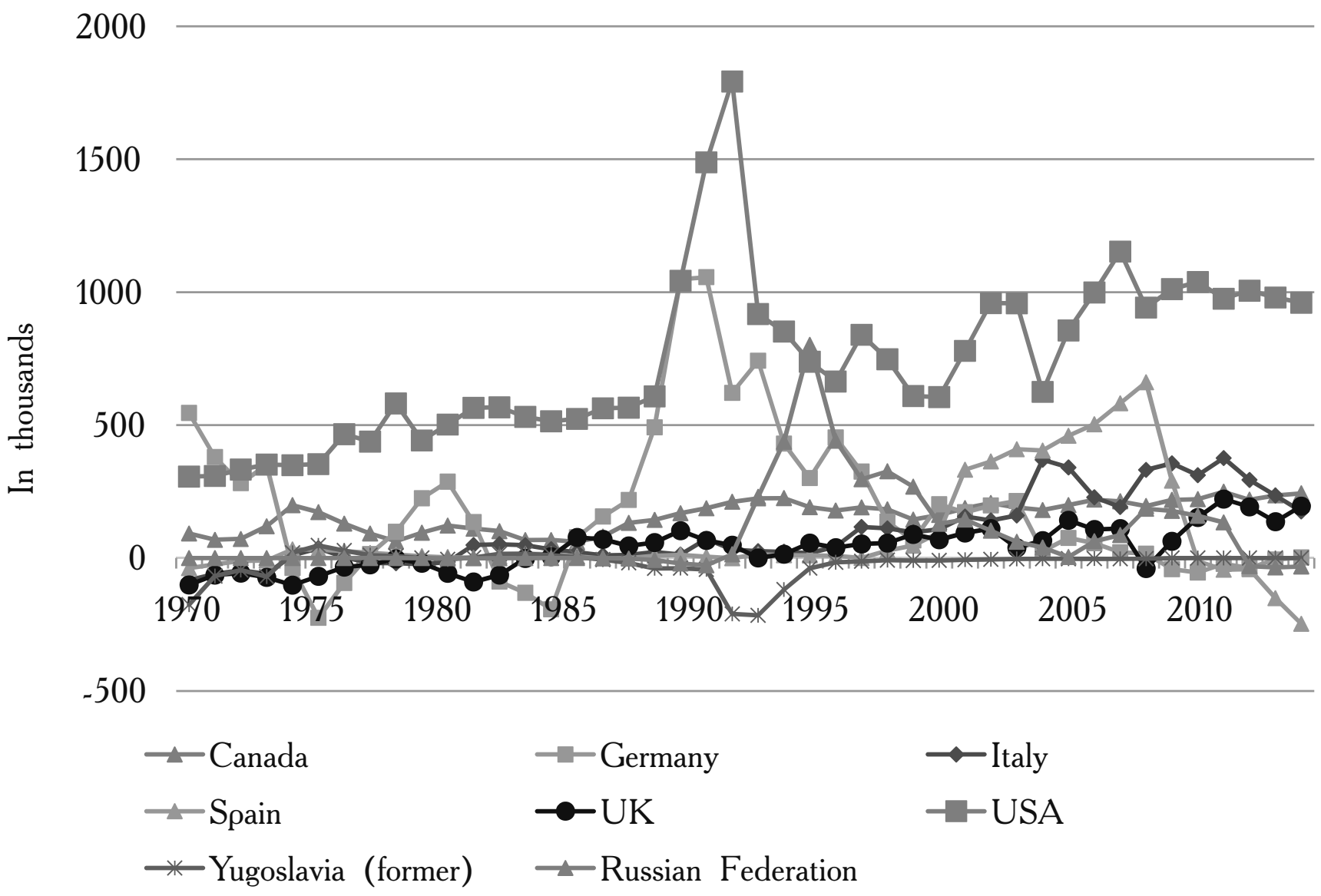

**Fig. 11.** Weighted Degree Difference (Net migration flow)

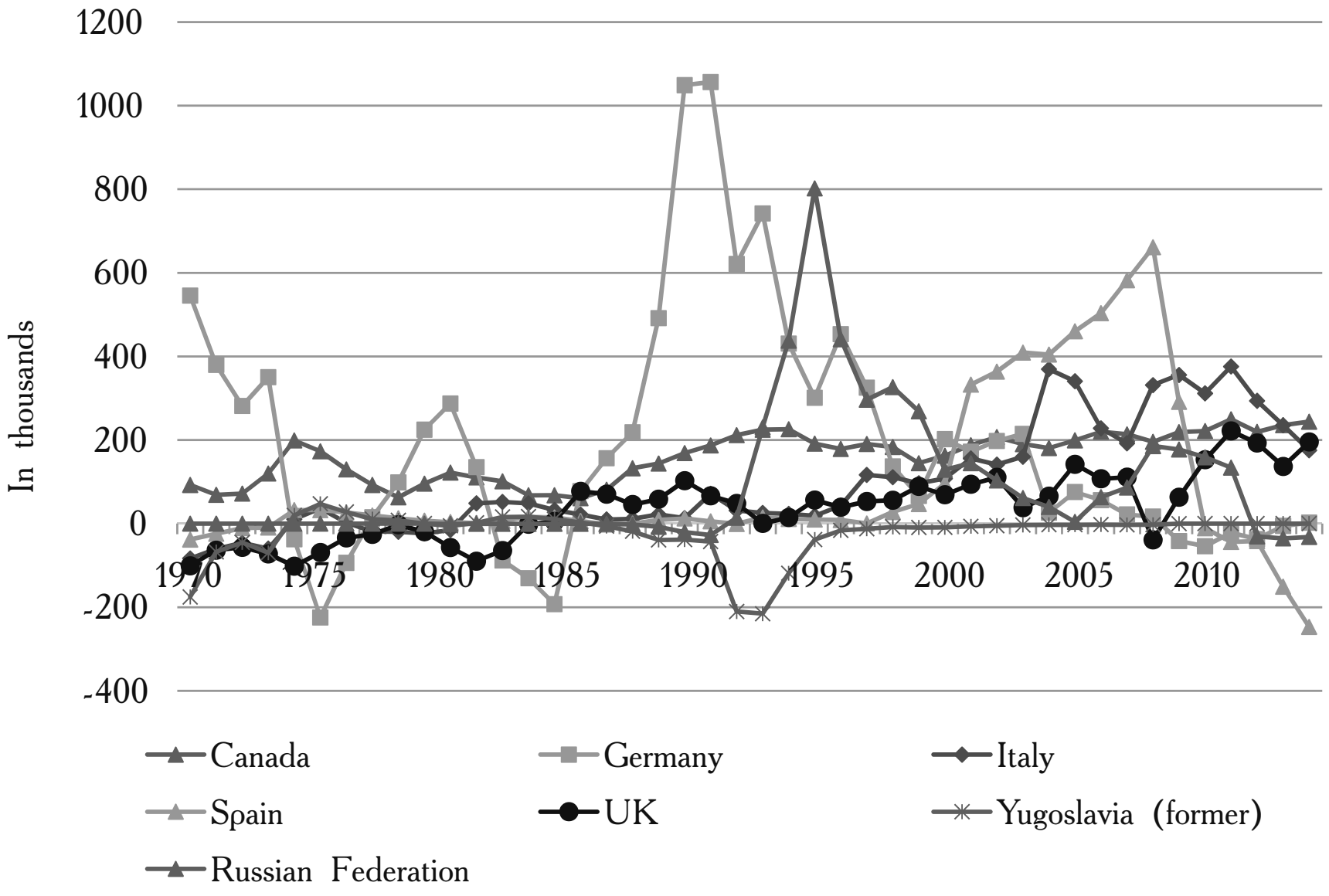

**Fig. 12.** Weighted Degree Difference (Net migration flow) without the USA

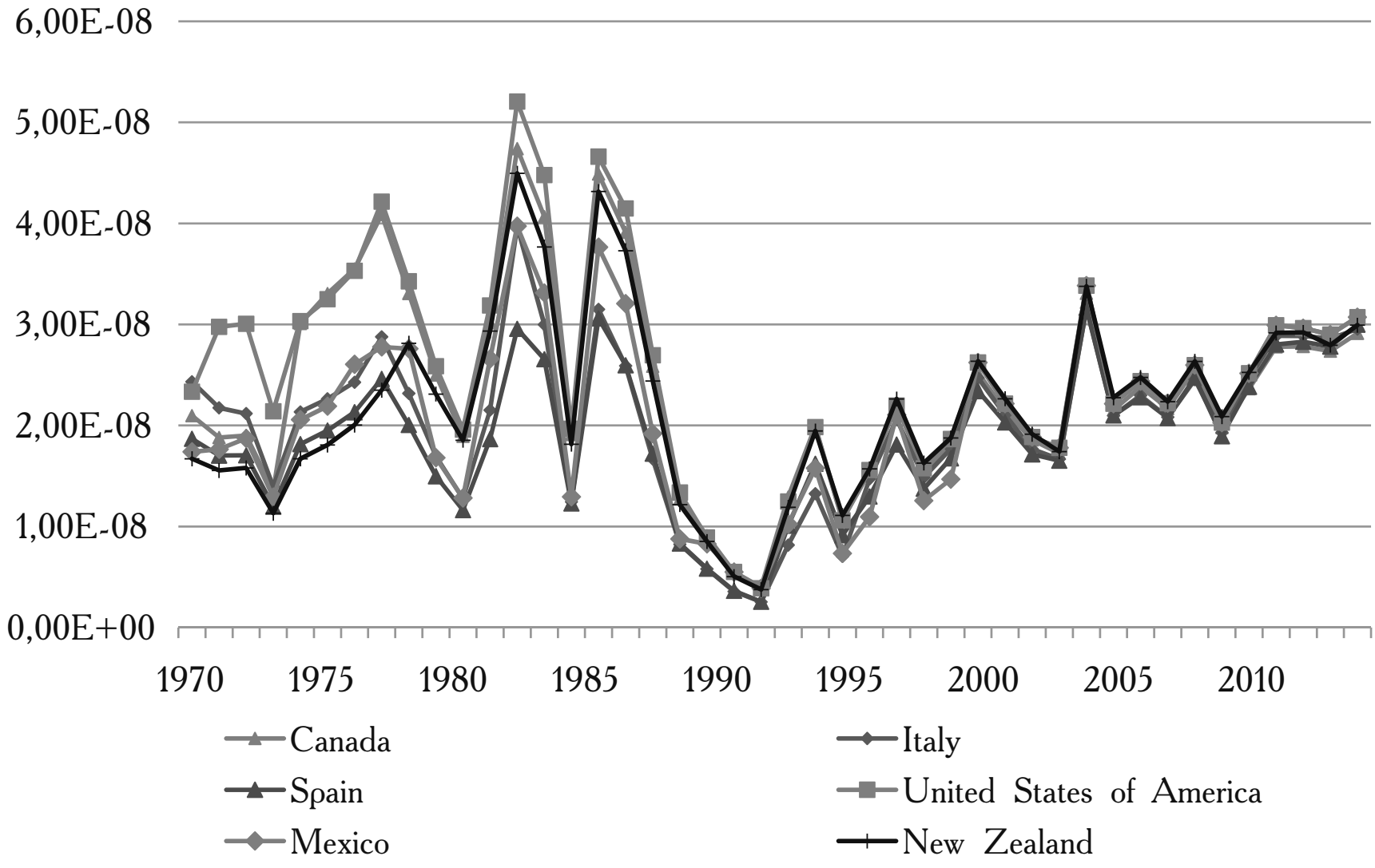

**Fig. 13.** Closeness

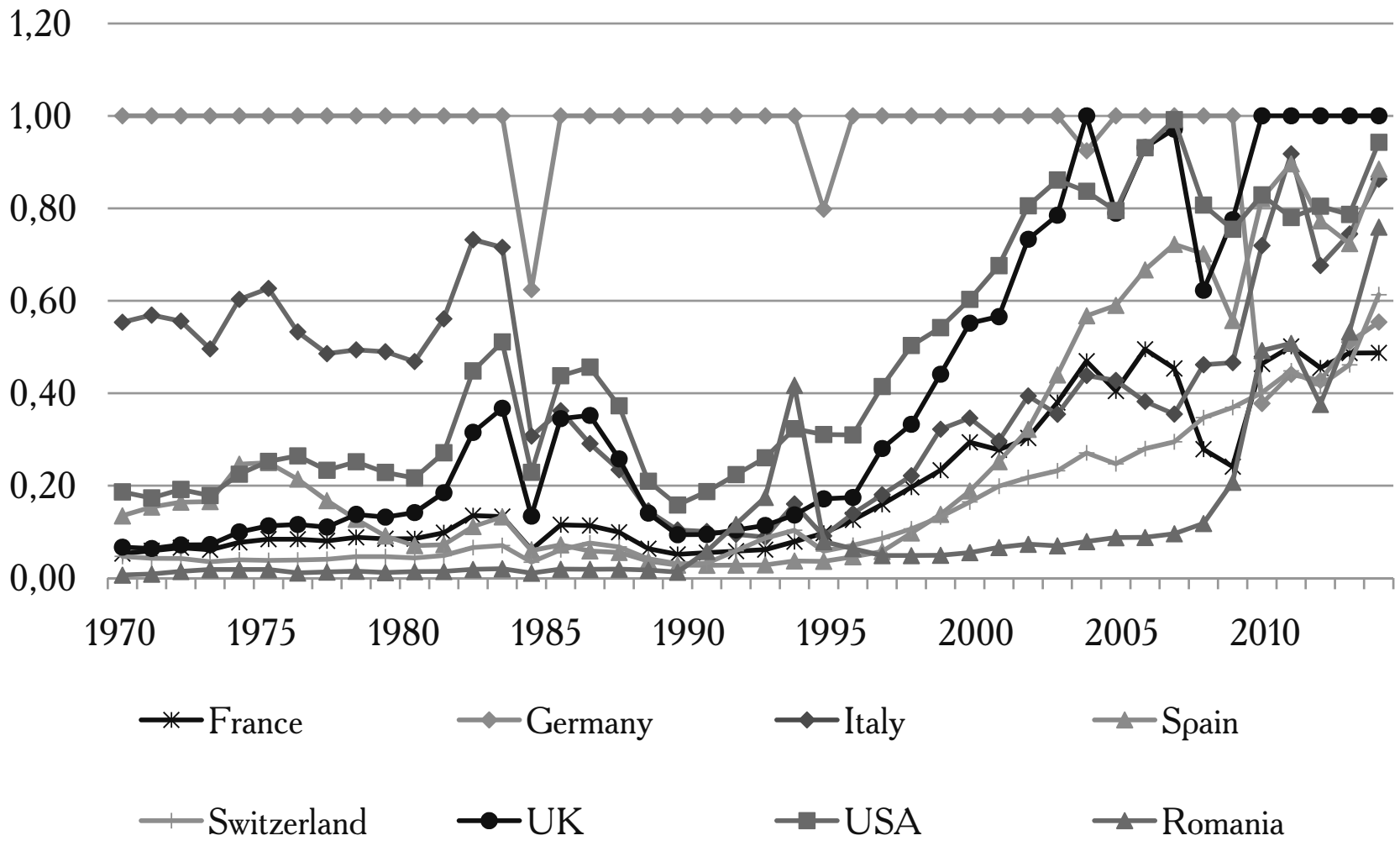

**Fig. 14.** Eigenvector

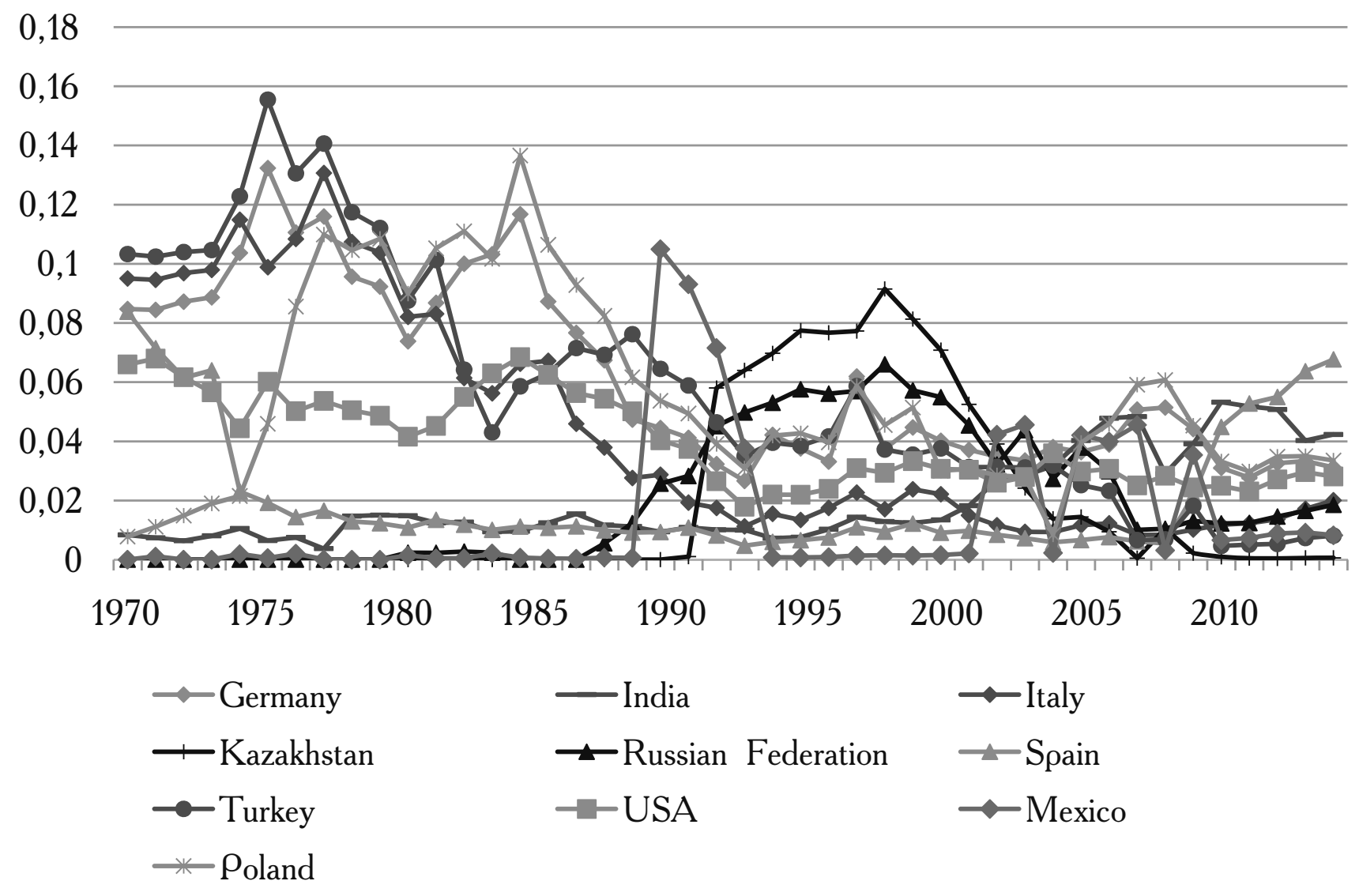

**Fig. 15.** LRIC (MAX)

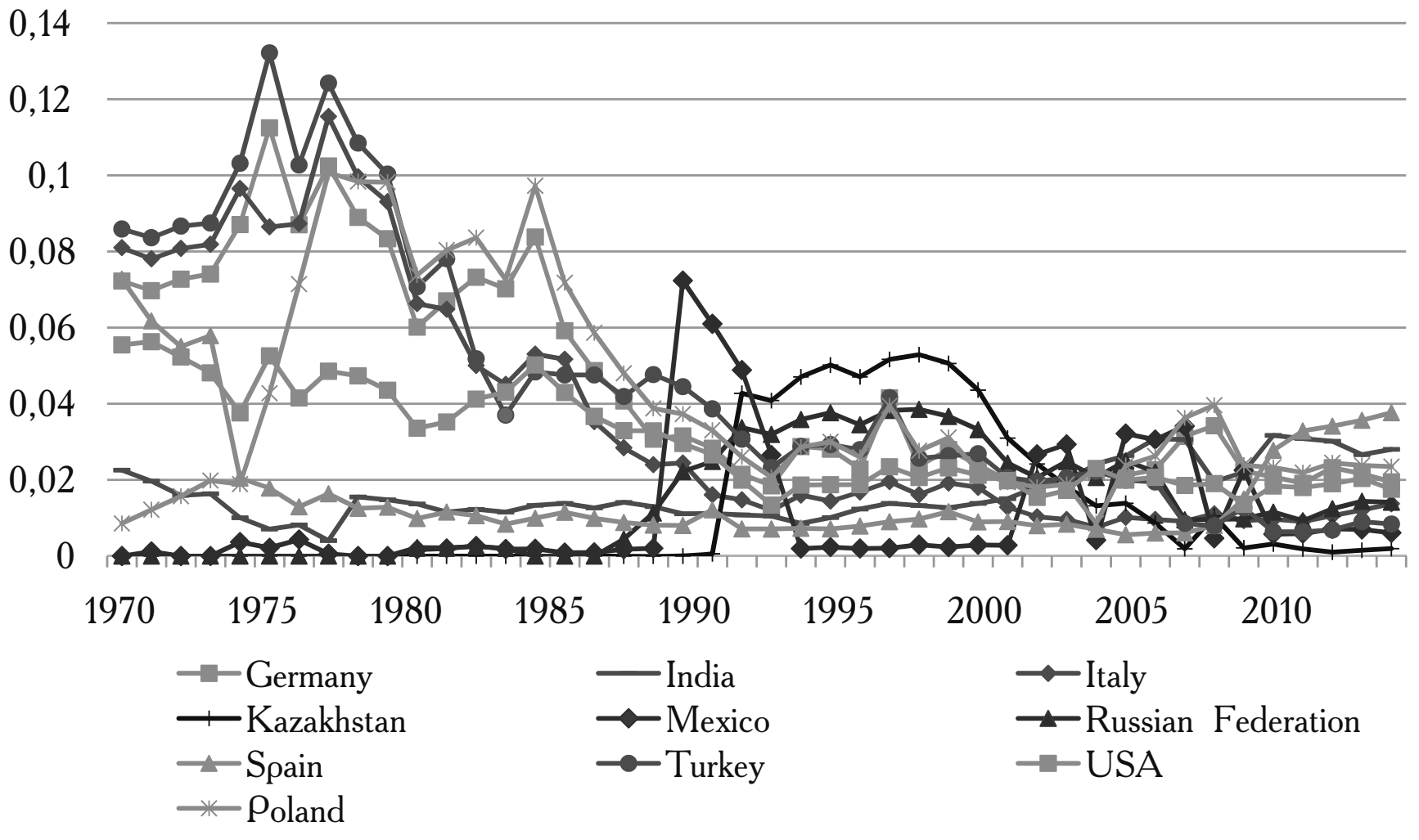

**Fig. 16.** LRIC (MAXMIN)

## References

[1] Aleskerov F.T. Power indices taking into account agents' preferences In: B. Simeone, F. Pukelsheim (eds), Mathematics and Democracy, Berlin: Springer, pp. 1–18, 2006.

[2] Aleskerov F., Andrievskaya I., Permjakova E. Key Borrowers Detected by the Intensities of Their Short-Range Interactions, Higher School of Economics Research Paper, 2014.

[3] Aleskerov F., Meshcheryakova N., Shvydun S. Centrality Measures in Networks based on Nodes Attributes, Long-Range Interactions and Group Influence. WP7/2016/04. Moscow: Higher School of Economics Publ. House, 2016 <https://www.hse.ru/en/org/hse/wp/wp7en>.

[4] Bavelas A. Communication patterns in task-oriented groups // J. Acoust. Soc. Am, 22 (6), 1950, 725–730.

[5] Bodvarsson Ö.B., Simpson N.B., Sparber C., 2015. Migration Theory // Handbook of economics of international migration. Vol. 1A The Immigrants (1st ed., Vol. 1). Elsevier Inc.

[6] Bonacich P. Power and centrality: A family of measures // American journal of sociology, 92 (5), 1987, pp. 1170–1182.

[7] Bonacich P. Technique for Analyzing Overlapping Memberships // Sociological Methodology, Vol. 4, 1972, pp. 176–185.

[8] Brin S., Page, L. The anatomy of a large-scale hypertextual Web search engine// Comput. Netw., 30, 1998, pp. 107–117.

[9] Brown, Richard P.C., Jimenez-Soto E., 2015. Migration and Remittances // Handbook of the Economics of International Migration (1st ed., Vol. 1B). Elsevier B.V.

[10] Davis K., D'Odorico P., Laio F., Ridolfi L., 2013. Global Spatio-Temporal Patterns in Human Migration: A Complex Network Perspective // PLoS NE8(1): e53723.

[11] Fagiolo G., Mastrorillo M., 2012. The International-Migration Network // eprint arXiv:1212.3852.

[12] Freeman L.C. A set of measures of centrality based upon betweenness // Sociometry, 40, pp. 35–41, 1977.

[13] Freeman L.C. Centrality in social networks: conceptual clarification // Social Networks, 1, pp. 215–239, 1979.

[14] Goodman L.A., Kruskal W.H. Measures of Association for Cross Classifications // Journal of the American Statistical Association Vol. 49. No. 268. P. 732–764, 1954.

[15] Izquierdo M., Jimeno J.F., Lacuesta A. Spain: from immigration to emigration? // Documentos de Trabajo, Banco de Espana, 2015.

[16] Katz L. A New Status Index Derived from Sociometric Index // Psychometrika, pp. 39–43, 1953.

[17] Lee E.A Theory of Migration // Demography, 3 (1), 47–57, 1966.

[18] Liang, Z., Li, J., & Ma, Z. Migration and Remittances // Asian Population Studies (1st ed., Vol. 9), 2013.

[19] Mincer J. Family migration decisions // J. Polit. Econ. 86 (5), 749–773, 1978.

[20] Polachek S.W., Horvath F.W. A Life Cycle Approach to Migration: Analysis of the Perspicacious Peregrinator // Research in Labor Economics 35th Anniversary Retrospective (Vol. 35, pp. 349–395), 2012.

[21] Ravenstein E.G., The Laws of Migration // Journal of the Royal Statistical Society, 52 (2), 241–305, 1889.

[22] Sjaastad L. The costs and returns of human migration // J. Polit. Econ. 70, 80–93, 1962.

[23] Smith A. An inquiry into the nature and causes of the wealth of nations (ed. RH Campbell, AS Skinner, and WB Todd), 1976.

[24] Tinbergen J. Shaping the world economy: suggestions for an international economic policy // № HD82 T54 (The Twentieth Century Fund, New York), 1962.

[25] Tranos E., Gheasi M., Nijkamp P. International migration: A global complex network. // Environment and Planning B: Planning and Design, 42 (1), 4–22, 2015.

[26] United Nations Development Program, 2009. Human Development Report – 2009 Overcoming barriers: Human mobility and development. Human Development (Vol. 331).

[27] United Nations, Department of Economic and Social Affairs, Population Division (2016). International Migration Report 2015: Highlights (ST/ESA/SER.A/375).

[28] United Nations. In safety and dignity: addressing large movements of refugees and migrants. Report of the Secretary-General, 21 April 2016

[29] Wanner P. Migration trends in Europe // European Population Papers Series No. 7, (7), 1–26, 2002.

[30] Yap L. The attraction of cities: A review of the migration literature // J. Dev. Econ. 4, 239–264, 1977.

[31] Zipf G.K. The P1 P2/d Hypothesis: On the Intercity Movement of Persons // American Sociological Review 11.6, 677–686, 1946.

**Internet sources**

[32] United Nations, Department of Economic and Social Affairs, Population Division (2009). International Migration Flows to and from Selected Countries: The 2008 Revision (United Nations database, POP/DB/MIG/Flow/Rev. 2008). Date of access: 30.11.2015.

[33] United Nations, Department of Economic and Social Affairs, Population Division (2015). International Migration Flows to and from Selected Countries: The 2015 Revision (POP/DB/MIG/Flow/Rev.2015). Date of access: 03.02.2016.

[34] World Bank, Global Bilateral Migration Database <http://data.worldbank.org/data-catalog/global-bilateral-migration-database> Date of access: 15.04.2016.

[35] MPI. Indian Immigrants in the United States <http://www.migrationpolicy.org/article/indian-immigrants-united-states> Date of access: 20.04.2016.

[36] MPI. Mexican Immigrants in the United States <http://www.migrationpolicy.org/article/mexican-immigrants-united-states> Date of access: 20.04.2016.

[37] Eurostat. Migration and Migrant population statistics <http://ec.europa.eu/eurostat/statistics-explained/index.php/Migration_and_migrant_population_statistics> Date of access: 28.05.2015.

[38] Migration Flows – Europe <http://migration.iom.int/europe/> Date of access: 15.04.2015.

Представлен анализ международных миграционных потоков с помощью применения сетевого анализа и расчета индексов центральности. С целью выявления наиболее влиятельных стран в сети международных потоков миграции рассмотрены классические индексы центральности и использованы новые индексы, учитывающие непрямые (ближние и дальние) взаимодействия вершин в сети, а также их индивидуальные атрибуты (информация о численности населения). Данная методология была применена для годовых данных по потокам международной миграции с 1970 по 2013 гг., предоставленных Организацией Объединенных Наций (ООН). Был проведен анализ результатов для одного года по каждому десятилетию, описана динамика. Показано, что классические индексы центральности и индексы ближних и дальних взаимодействий выделяют страны с большим притоком/оттоком мигрантов, а индексы ближних и дальних взаимодействий выделяют страны со значительным оттоком мигрантов в страны с наибольшим притоком мигрантов и наиболее взаимосвязанные с ними страны в сети международной миграции.

*Алескеров Ф.Т.*, Национальный исследовательский университет «Высшая школа экономики» (НИУ ВШЭ), Международная научно-учебная лаборатория анализа и выбора решений, Институт проблем управления им. В.А. Трапезникова Российской академии наук (ИПУ РАН), Москва; alesk@hse.ru

*Мещерякова Н.Г.*, Национальный исследовательский университет «Высшая школа экономики» (НИУ ВШЭ), Международная научно-учебная лаборатория анализа и выбора решений, Институт проблем управления им. В.А. Трапезникова Российской академии наук (ИПУ РАН), Москва; natamesc@gmail.com

*Резяпова А.Н.*, Национальный исследовательский университет «Высшая школа экономики» (НИУ ВШЭ), Международная научно-учебная лаборатория анализа и выбора решений, Москва; annrezyapova@gmail.com

*Швыдун С.В.*, Национальный исследовательский университет «Высшая школа экономики» (НИУ ВШЭ), Международная научно-учебная лаборатория анализа и выбора решений, Институт проблем управления им. В.А. Трапезникова Российской академии наук (ИПУ РАН), Москва; shvydun@hse.ru

Алескеров Фуад, Мещерякова Наталия,
Резяпова Анна, Швыдун Сергей

# **Сетевой анализ международных миграционных потоков**

(*на английском языке*)